\documentclass[prd,onecolumn,nopacs,nofootinbib]{revtex4}

\usepackage{amsmath}
\usepackage{graphicx}
\usepackage{amsfonts}
\usepackage{latexsym}
\usepackage{bbold}
\usepackage{wasysym}
\usepackage{calligra}
\usepackage{float}
\usepackage{ulem}
\usepackage{inputenc}
\usepackage{xspace}
\usepackage{url}
\usepackage{epstopdf}
\usepackage{tikz}

\newcommand\Lie{\pounds}

\newtheorem{prop}{Proposition}

\newcommand{\be}{\begin{equation}}
\newcommand{\ee}{\end{equation}}
\newcommand{\beq} {\begin{equation}}
\newcommand{\eeq} {\end{equation}}
\newcommand{\ba}{\begin{eqnarray}}
\newcommand{\ea}{\end{eqnarray}}

\begin{document}

	\title{Cosmological Hyperfluids, Torsion and Non-metricity}
	
\author{Damianos Iosifidis}
\affiliation{Institute of Theoretical Physics, Department of Physics
	Aristotle University of Thessaloniki, 54124 Thessaloniki, Greece}
\email{diosifid@auth.gr}
	
	\date{\today}
	\begin{abstract}
		We develop a novel model for Cosmological Hyperfluids, that is fluids with intrinsic hypermomentum that induce spacetime torsion and non-metricity.
		Imposing the Cosmological Principle to Metric-Affine Spaces, we present the most general covariant form of the hypermomentum tensor in an FLRW Universe along with its conservation laws and therefore construct a novel hyperfluid model for Cosmological purposes. Extending the previous model of the unconstrained hyperfluid in a Cosmological setting we establish the conservation laws for energy-momentum and hypermomentum and therefore provide the complete Cosmological setup to study non-Riemannian effects in Cosmology. With the help of this we find the forms of torsion and non-metricity that were earlier reported in the literature and also obtain the most general form of the Friedmann equations with torsion and non-metricity. We also discuss some applications of our model, make contact with the known results in the literature and point to future directions.
		
	\end{abstract}
	
	\maketitle
	
	\allowdisplaybreaks
	
	
	\tableofcontents
	
	\section{Introduction}
	\label{intro}
	 General Relativity is undoubtedly one of the most well established, mathematically beautiful and properly formulated Theories of Physics. Its nice geometrical interpretation along with its solid predictions give enough reasons to call the latter a successful Theory. 
	However, despite its great success General Relativity falls short in explaining the current Cosmological data. More specifically, it fails to properly explain the Cosmological evolution at early times and cannot predict a late time accelerated expansion. As a result many Theories of Modified Gravity have been proposed\cite{clifton2012modified}. Among the many possibilities, one most intriguing, well motivated and in the spirit of geometrization of Gravity is to  generalize the affine connection and relax the torsion-free and metric-compatibility conditions. Then one is dealing with a non-Riemannian Geometry\cite{eisenhart2012non,schouten2013ricci}, where curvature torsion and non-metricity are the intrinsic characteristics of space. Restricting the aforementioned non-Riemannian geometry in a certain way one obtains different formulations of Gravity. For instance, imposing vanishing torsion and non-metricity we arrive at the well known metric Theories, of which GR is a special case. On the other hand, demanding vanishing curvature and non-metricity we have the standard teleparallel formulation\cite{aldrovandi2012teleparallel} while if we allow only for non-metricity and set curvature and torsion to zero we get the symmetric teleparallel scheme\cite{nester1998symmetric,jimenez2018teleparallel}. Yet, it is also possible to impose only vanishing curvature and arrive at a generalized teleparallel formulation exhibiting  both torsion and no-metricity\cite{jimenez2019general}. Leaving the geometrical objects of the space free of constraints, we have the most general Gravitational Theory supporting a non-Riemannian Geometry, which is known as Metric-Affine Gravity(MAG)\cite{hehl1995metric,hehl1999metric,iosifidis2019metric}.

	In this article we will focus on the latter.  MAG is unique in its own right, it is known to be a Gauge Theory of Gravity\cite{hehl1995metric}, it takes into account the microscopic characteristics of matter (see discussion bellow) and also the modifications it introduces  are of purely geometrical nature. Indeed as we mentioned, the geometric arena of MAG is a generalized non-Riemannian Geometry where apart from curvature, the space also possesses torsion and non-metricity. That is, the extra degrees of freedom (compared to GR) are due to the torsion and non-metricity of spacetime which are linked to the micro-structure of matter. From the above discussion becomes clear that fluids that carry hypermomentum are very interesting especially with regards to their Cosmological applications. In this direction, there are some very interesting models of matter that carries hypermomentum, these include the ideal hyperfluid\cite{obukhov1993hyperfluid} and the unconstrained hyperfluid\cite{obukhov1996model}. The first model\cite{obukhov1993hyperfluid} is actually a generalization of the so-called Weyssenhoff fluid\cite{weyssenhoff1947relativistic} for spinning particles where now the fluid has also non-metric degrees of freedom but it continuous to satisfy the Frenkel-type condition $J_{\mu\nu}u^{\mu}=0=J_{\mu\nu}u^{\nu}$, where $J_{\mu\nu}$ is the hypermomentum density. That is why this model is dubbed the constrained hyperfluid. Then, it was suggested in \cite{babourova1995variational} that the hyperfluid should be free from the Frenkel constraint and in \cite{obukhov1996model} Obukhov presented the very interesting model of the unconstrained hyperfluid\footnote{An alternative and also interesting model for perfect hyperfluid was developed in \cite{babourova1998perfect}. Yet another interesting model of a fluid possessing both spin and twist was constructed in \cite{smalley1995fluids}. }. We will briefly review the latter in the subsequent chapters and also discuss how it can be used to obtain some (but not all) of the non-Riemannain degrees of freedom in Cosmology. We will then propose an extension which, as we will show, produces all the torsional and non-metric degrees of freedom that are present in Cosmology. Our proposal for the Perfect Cosmological Hyperfluid follows directly by applying the Cosmological Principle to the Hypermomentum tensor and the conservation laws associated with it are obtained, as usual, by the diffeomorphism invariance of the matter sector of the action. 
 	
	The article is organized as follows. Firstly, we make a brief introduction in non-Riemannian geometry and introduce our conventions etc. We then turn our attention on the hypermomentum and energy momentum (canonical and metrical) tensors and also briefly review the model of the unconstrained hyperfluid. Continuing we discuss the non-Riemannian degrees of freedom in an FLRW Cosmology and explain how the above hyperfluid model can produce but few of the aforementioned degrees of freedom. We then present our novel model for the Perfect (or Ideal) Cosmological Hyperfluid. By extending the Cosmological Principle to the hypermomentum, we impose a vanishing Lie derivative for the latter and obtain its most general covariant form in a homogeneous Universe, which is spanned by $5$ degrees of freedom. Then decomposing the hypermomentum, we show how these $5$ degrees of freedom nicely split into $2+3$, two for torsion and three for non-metricity respectively. We then, equip our model with the conservation laws for the energy momentum and hypermomentum which follow by the diffeomorphism invariance of the matter action and are essentially the same with the case of the unconstrained hyperfluid. Having obtained the complete Cosmological setup for the Perfect Cosmological Hyperfluid, we consider a MAG model consisting of the usual Einstein-Hilbert action and consider the matter sector to be that of the Perfect Cosmological Hyperfluid. First, considering a vanishing non-metricity and then a vanishing torsion, we obtain the most straightforward modifications of the Friedmann equations in the presence of solely torsion and non-metricity respectively. Then allowing for both we derive the modified Friedmann equations in a general non-Riemannian Geometry (i.e. both torsion and non-metricity non-vanishing) and also discuss some subtleties associated with the projective invariance of the gravitational action and the constraints it imposes on the hyperfluid. We wrap up our results and discuss applications of the proposed hyperfluid model.
	\section{Non-Riemannian Geometry}
In this section we briefly review the basic aspects of a non-Riemmannian geometry and set up the definitions for the various geometrical objects we are going to be using throughout.

The structure of a non-Riemannian space is determined by the metric tensor $g_{\mu\nu}$ which measures distances, defines dot products and raises and lowers indices along with an affine connection $\Gamma^{\lambda}_{\;\;\;\mu\nu}$ (or just $\nabla$) which defines parallel transfer of tensor fields through covariant differentiation. In our conventions the covariant derivative of, say, a $(1,1)$ type tensor reads
\beq
\nabla_{\alpha}T^{\mu}_{\;\;\nu}=\partial_{\alpha}T^{\mu}_{\;\;\nu}-\Gamma^{\lambda}_{\;\;\;\nu\alpha}T^{\mu}_{\;\;\lambda}+\Gamma^{\mu}_{\;\;\;\lambda\alpha}T^{\lambda}_{\;\;\nu}
\eeq
The antisymmetric part of the affine connection defines the torsion tensor
\beq
S_{\mu\nu}^{\;\;\;\lambda}:=\Gamma^{\lambda}_{\;\;\;[\mu\nu]}
\eeq
which naturally arises by acting the antisymmetrized covariant derivative on a scalar
\beq
\nabla_{[\mu}\nabla_{\nu]}\phi=S_{\mu\nu}^{\;\;\;\lambda}\nabla_{\lambda}\phi
\eeq
Given a torsion tensor  contracting it in its last two indices we may define the torsion vector
\beq
S_{\mu}:=S_{\mu\lambda}^{\;\;\;\;\lambda}
\eeq
which is obtained without the use of any metric and exists for arbitrary space dimension-$n$. For $n=4$ in particular there also exists the torsion pseudo-vector
\beq
\hat{S}_{\mu}:=\epsilon_{\mu\alpha\beta\gamma}S^{\alpha\beta\gamma}
\eeq
where  $\epsilon_{\mu\alpha\beta\gamma}$ is the $d-dimensional$ totally antisymmetric Levi-Civita tensor. 
Considering again the antisymmetrized covariant derivative, but now acting it on a vector field $u^{\mu}$ it follows that
\begin{equation}
[\nabla_{\alpha} ,\nabla_{\beta}]u^{\mu}=2\nabla_{[\alpha} \nabla_{\beta]}u^{\mu}=R^{\mu}_{\;\;\;\nu\alpha\beta} u^{\nu}+2 S_{\alpha\beta}^{\;\;\;\;\;\nu}\nabla_{\nu}u^{\mu}
\end{equation}
where
\begin{equation}
R^{\mu}_{\;\;\;\nu\alpha\beta}:= 2\partial_{[\alpha}\Gamma^{\mu}_{\;\;\;|\nu|\beta]}+2\Gamma^{\mu}_{\;\;\;\rho[\alpha}\Gamma^{\rho}_{\;\;\;|\nu|\beta]}
\end{equation}
is the so-called Riemann or Curvature tensor and the horizontal bars around an index denote that this index is left out of the (anti)-symmetrization. In non-Riemannian Geometries the only symmetry the latter possesses is antisymmetry in its last two indices as it is obvious by its definition above.  Without the use of any metric, we can construct the  two independent contractions for the Riemann tensor
\beq
R_{\nu\beta}:=R^{\mu}_{\;\;\;\nu\mu\beta}\;,\; \hat{R}_{\alpha\beta}:=R^{\mu}_{\;\;\;\mu\alpha\beta}
\eeq
The former defines as usual the Ricci tensor while the latter is the tensor of homothetic curvature and is of purely non-Riemannian origin. Once a metric is given we can form yet another  contraction
\beq
\check{R}^{\lambda}_{\;\;\kappa}:=R^{\lambda}_{\;\;\mu\nu\kappa}g^{\mu\nu}
\eeq
However, the Ricci scalar is still uniquely defined since
\beq
R:=R_{\mu\nu}g^{\mu\nu}=-\check{R}_{\mu\nu}g^{\mu\nu}\;,\;\; \hat{R}_{\mu\nu}g^{\mu\nu}=0
\eeq

Now, in a generic non-Riemannian space the metric need not be covariantly conserved, and we say that the connection is not metric-compatible. Exactly this deviation from  compatibility, defines the non-metricity tensor
\beq
Q_{\alpha\mu\nu}=-\nabla_{\alpha}g_{\mu\nu}
\eeq
The above can be contracted in two independent ways, giving us the two non-metricity vectors
\beq 
Q_{\alpha}:=Q_{\alpha\mu\nu}g^{\mu\nu}\;,\;\; \hat{Q}_{\nu}=Q_{\alpha\mu\nu}g^{\alpha\mu}
\eeq
The former is commonly referred to as the Weyl vector. With the above definitions it is trivial to show (see for instance \cite{schouten2013ricci,iosifidis2019metric}) that the general affine-connection can be decomposed according to
\beq
\Gamma^{\lambda}_{\;\;\;\mu\nu}=\tilde{\Gamma}^{\lambda}_{\;\;\;\mu\nu}+N^{\lambda}_{\;\;\;\mu\nu} \label{condec}
\eeq
where 
	\beq
N_{\alpha\mu\nu}=\frac{1}{2}(Q_{\mu\nu\alpha}+Q_{\nu\alpha\mu}-Q_{\alpha\mu\nu}) -(S_{\alpha\mu\nu}+S_{\alpha\nu\mu}-S_{\mu\nu\alpha}) \label{l}
\eeq
is called the distortion tensor and $\tilde{\Gamma}^{\lambda}_{\;\;\;\mu\nu}$ is the usual Levi-Civita connection given by
\begin{equation}
\tilde{\Gamma}^{\lambda}_{\;\;\;\mu\nu}:=\frac{1}{2}g^{\alpha\lambda}(\partial_{\mu}g_{\nu\alpha}+\partial_{\nu}g_{\alpha\mu}-\partial_{\alpha}g_{\mu\nu})
\end{equation}
Objects appearing with a  tilde will always denote Riemannian parts unless otherwise stated. With the above connection decomposition we can express any geometrical object into its Riemannian piece plus non-Riemannian contributions coming from torsion and non-metricity. For instance, substituting the decomposition ($\ref{condec}$) into the definition of the Riemann tensor we deduce
\beq
R^{\mu}_{\;\;\nu\alpha\beta}=\tilde{R}^{\mu}_{\;\;\nu\alpha\beta} +2\tilde{\nabla}_{[\alpha}N^{\mu}_{\;\;\;|\nu|\beta]}
+2 N^{\mu}_{\;\;\;\lambda [\alpha}N^{\lambda}_{\;\;\;|\nu|\beta]}
\eeq
and by contracting the latter we also get the decompositions for the Ricci tensor, Ricci scalar etc. Having developed the minimum  geometrical setup needed for the rest of our analysis will now turn our attention to Metric-Affine Gravity(MAG).

	\section{ Hypermomentum, Canonical and Metrical Energy Momentum Tensors}
	As it is known in the general Metric-Affine Gravity (MAG) framework the action is a functional of the metric, the independent affine connection and the matter fields, namely\footnote{Of course the action will also depend on the derivatives of the metric and the connection. We are suppressing this dependence on the grounds of convenience.}
	\beq	
	S[g,\Gamma,\phi]=S_{G}[g,\Gamma]+S_{M}[g,\Gamma,\phi]
	\eeq
	where
	\beq
	S_{G}[g,\Gamma]=\frac{1}{2 \kappa}\int d^{n}x \sqrt{-g} \mathcal{L}_{G}(g,\Gamma)
	\eeq
		\beq
	S_{M}[g,\Gamma,\phi]=\int d^{n}x \sqrt{-g} \mathcal{L}_{M}(g,\Gamma,\phi)
	\eeq
	represent the gravitational and matter sectors respectively and $\phi$ collectively denotes the matter fields.	Now, given a matter action, we define as usual the metrical (symmetric) energy-momentum tensor(MEMT) by
	\beq
T_{\alpha\beta}:= -\frac{2}{\sqrt{-g}}\frac{\delta S_{M}}{\delta g^{\alpha\beta}}=-\frac{2}{\sqrt{-g}}\frac{\delta(\sqrt{-g} \mathcal{L}_{M})}{\delta g^{\alpha\beta}}
\eeq
	In MAG, the connection naturally couples to matter so we now also have the variation
\beq
\Delta_{\lambda}^{\;\;\;\mu\nu}:= -\frac{2}{\sqrt{-g}}\frac{\delta S_{M}}{\delta \Gamma^{\lambda}_{\;\;\;\mu\nu}}=-\frac{2}{\sqrt{-g}}\frac{\delta ( \sqrt{-g} \mathcal{L}_{M})}{\delta \Gamma^{\lambda}_{\;\;\;\mu\nu}}
\eeq
	which is the so-hypermomentum tensor(HMT)\cite{hehl1976hypermomentum} and encompasses the microscopic characteristics of matter such as spin, dilation and shear\cite{hehl1995metric}. Note that if one works in the equivalent  formalism based on the vielbeins $e_{\mu}^{\;\; c}$\footnote{Here Latin indices $a,b,c...$ denote Lorentz indices, that is tangent space indices. The connection between the metric and the vielbeins is, as usual, $g_{\mu\nu}=e_{\mu}^{\;\;a}e_{\nu}^{\;\;b} \eta_{ab}$ where $\eta_{ab}$ is the tangent space (flat) Minkowski metric tensor.} and the spin connection $\omega_{\mu a b}$ then we have the definition of the canonical energy-momentum tensor(CEMT) given by
	\beq
	t^{\mu}_{\;\; c}=\frac{1}{\sqrt{-g}}\frac{\delta S_{M}}{\delta e_{\mu}^{\;\; c}}
	\eeq
	which is not symmetric in general. We should stress out that the latter is not independent from the above defined MEMT and HMT\footnote{This is also apparent if one regards, $e_{\mu}^{\;\; c},g_{\mu\nu}$ and $\Gamma^{\lambda}_{\;\;\;\mu\nu}$ as independent variables and start with an action $S[e,g,\Gamma]$. It can be shown then, that the field equations for the metric tensor are not independent from  the set coming from the $e$ and $\Gamma$ variations\cite{hehl1995metric}. }. Indeed, from its very definition, the use of the chain rule and converting all tangent space indices to holonomic ones we find
	\beq
	t^{\mu}_{\;\;\lambda}:=\frac{1}{\sqrt{-g}}\frac{\delta S_{M}}{\delta e_{\mu}^{\;\; c}}e_{\lambda}^{\;\;c}= T^{\mu}_{\;\;\lambda}-\frac{1}{2 \sqrt{-g}}\hat{\nabla}_{\nu}(\sqrt{-g}\Delta_{\lambda}^{\;\;\mu\nu}) \label{EMTs}
	\eeq
	where 
	\beq
	\hat{\nabla}_{\nu}=2S_{\nu}-\nabla_{\nu}
	\eeq
	is the modified covariant derivative and	we have also made use of the identity
	\begin{equation}
	\nabla_{\nu}e_{\mu}^{\;\; a}=0=\partial_{\nu}e_{\mu}^{\;\; a}-\Gamma^{\rho}_{\;\;\;\mu\nu}e_{\rho}^{\;\;a}+\omega_{\nu\;\;b}^{\;\;a}e_{\mu}^{\;\;b}
	\end{equation}
	that connects the two formalisms and is oftentimes called the vielbein postulate (even though it is not!).
	 Of course, this connection was known in the literature a long time ago (see for instance  \cite{babourova1995peculiarities,puetzfeld2008probing}). Equation $(\ref{EMTs})$ will be our reference point in constructing the Cosmological Hyperfluid Model. In fact, as we will demonstrate this very equation would stem as the starting point for the first conservation law of the Cosmological Hyperfluid.  Note now that bringing all indices down in($\ref{EMTs}$) and antisymmetrizing one arrives at the conservation law for spin\cite{obukhov2014conservation}
	\beq
	2t_{[\mu\nu]}=\frac{1}{\sqrt{-g}}\hat{\nabla}_{\alpha}(\sqrt{-g}\tau_{\mu\nu}^{\;\;\;\;\alpha})-Q_{\alpha\beta[\mu}\Delta_{\nu]}^{\;\;\beta\alpha}
	\eeq
	\beq
	\tau_{\mu\nu}^{\;\;\;\;\alpha}:=\Delta_{[\mu\nu]}^{\;\;\;\;\;\alpha}
	\eeq
	It is  worth mentioning that the conservation law for spin receives, in general, contributions from non-metricity as is obvious from the above exposure. It is also evident from ($\ref{EMTs}$) that for matter with no microstructure ($\Delta_{\alpha\mu\nu}\equiv 0$) the canonical and the metrical energy-momentum tensors coincide. In addition contracting ($\ref{EMTs}$) in $\mu=\lambda$ we get the trace relation
	\beq
	t=T+\frac{1}{2 \sqrt{-g}}\partial_{\nu}(\sqrt{-g}\Delta^{\nu})
	\eeq
	where
	\beq
	t:=t^{\mu}_{\;\;\mu}\;\;,\;\; T:=T^{\mu}_{\;\;\mu}\;\;,\;\; \Delta^{\nu}:=\Delta_{\lambda}^{\;\;\lambda\nu}
	\eeq
	 and we have used the trivial identity $\hat{\nabla}_{\nu}\mathcal{D}^{\nu}=-\partial_{\nu}\mathcal{D}^{\nu}$ which holds for arbitrary vector densities of weight $w=1$. From the last equation we see that for specific matter types,  the following relations hold true 
	 \beq
	 T=0 \;\;\; \Leftrightarrow \;\;\; 2t=\frac{1}{ \sqrt{-g}}\partial_{\nu}(\sqrt{-g}\Delta^{\nu})
	 \eeq
	  \beq
	 t=0 \;\;\; \Leftrightarrow \;\;\; 2 T=-\frac{1}{ \sqrt{-g}}\partial_{\nu}(\sqrt{-g}\Delta^{\nu}) \label{tTT}
	 \eeq
	 \beq
		 t=T \;\;\; \Leftrightarrow \;\;\;\partial_{\nu}(\sqrt{-g}\Delta^{\nu})=0
	 \eeq
	 
	 corresponding to the cases of a conformally invariant, a frame rescaling invariant and  special projective transformation invariant Theories respectively (see also \cite{iosifidis2019scale}).

	\section{Unconstrained Hyperfluid}
	As we have discussed, MAG takes into account the microscopic characteristics of matter such as spin, dilation and shear\cite{hehl1995metric}. These properties are encoded in the  hypermomentum tensor and result in the creation of spacetime torsion and non-metricity. Perhaps the most known example of a fluid with micro-structure is the Weyssenhoff\cite{weyssenhoff1947relativistic} spinning fluid which produces a non-vanishing spacetime torsion and fits nicely into the  scheme of Einstein-Cartan Theory. However, such a fluid is not compatible with the Cosmological Principle, as shown by Tsamparlis\cite{tsamparlis1979cosmological}, and therefore this kind of torsion cannot exist in a homogeneous FLRW Universe. There are however two torsional degrees of freedom that are allowed in a homogeneous $4-dim$ Cosmology, one of which was analyzed in \cite{kranas2019friedmann}(see also \cite{tsamparlis1981methods} for the general case). The task would now be to develop a fluid model that encompasses the non-metric degrees of freedom as well. Such models do exist and are the  ideal (constrained) hyperfluid\cite{obukhov1993hyperfluid} and the revised unconstrained hyperfluid\cite{obukhov1996model}, the latter being a refined version of the former where the Frenkel-type condition is relaxed. However, as we will demonstrate by extending the Cosmological Principle, this model can produce only two of the five known non-Riemannian degrees of freedom in an FLRW Universe and, in addition, cannot account for torsional degrees of freedom. By extending it in a certain way so as to respect the Cosmological Principle we will present the most general covariant form of the latter that produces the expected non-Riemannian Cosmological degrees of freedom.  
	Before doing so,  
	 let us first briefly review the unconstrained hyperfluid model which is closer to the Cosmological Hyperfluid we are proposing here and will give us an insight on how to proceed. First, we note that for the unconstrained hyperfluid, the metric stress-energy and the canonical energy momentum tensor coincide\cite{obukhov1996model} and the conservation laws for energy-momentum and hypermomentum read
\beq
-\frac{2}{\sqrt{-g}}\hat{{\nabla}}_{\mu}( \sqrt{-g}T^{\mu}_{\;\;\nu})=\Delta^{\alpha\beta\gamma}R_{\alpha\beta\gamma\nu}-\frac{1}{2}Q_{\nu\alpha\beta}T^{\alpha\beta}-2S_{\nu\alpha\beta}T^{\alpha\beta} \label{Tcons1}
\eeq
	\beq
	\hat{\nabla}_{\nu}\Big( \sqrt{-g}\Delta_{\lambda}^{\;\;\;\mu\nu}\Big)=0 \label{DD}
	\eeq
	which, as usual, are obtained from the diffeomorphism invariance of the matter action.
	In addition, in  this case the forms of $T_{\mu\nu}$ and $\Delta_{\alpha\mu\nu}$ decouple and are given by
	\beq
	T_{\mu\nu}=\rho u_{\mu} u_{\nu}+p h_{\mu\nu} \label{Tform}
	\eeq
	\beq
	\Delta_{\lambda}^{\;\;\;\mu\nu}=J_{\lambda}^{\;\;\mu}u^{\nu} \label{dJ}
	\eeq
	respectively\cite{obukhov1996model}, in contrast to the (constrained) hyperfluid where their forms mix-up\cite{obukhov1993hyperfluid}. In the above, $T_{\mu\nu}$ is the usual energy-momentum of the perfect fluid and $J_{\mu\nu}$ is the hypermomentum density of the hyperfluid. In the first hyperfluid model, the latter obeyed the Frenkel-type condition
	\beq
	J_{\mu\nu}u^{\mu}=0=J_{\mu\nu}u^{\nu}
	\eeq 
	which is a direct generalization of the Frenkel condition on spin density $s_{\mu\nu}u^{\mu}=0$ ($s_{\mu\nu}=-s_{\nu\mu}$ being the spin density). Note that when non-metricity is switched-off the hypermomentum density coincides with the spin density (i.e. $J_{\mu\nu}=s_{\mu\nu}$). In the general case, the spin density is given by the antisymmetric part of $J_{\mu\nu}$ which contains a symmetric part as well. More information regarding the constrained and the unconstrained ideal hyperfluid may be found in the original and most interesting papers of Obukov\cite{obukhov1993hyperfluid,obukhov1996model}. We shall now discuss how the above model is applied to Cosmology and also propose a slight generalization in order to establish what we shall call  a Perfect (or Ideal) Cosmological Hyperfluid. Before doing so we will briefly  discuss the role of torsion and non-metricity in Cosmology. 
	
	\section{Cosmology with torsion and Non-metricity}
	To begin with let us note that in a Cosmological setting, torsion has one or two degrees of freedom depending on the spatial dimensionality. Indeed, as shown by Tsamparlis\cite{tsamparlis1979cosmological} if we consider a $1+(n-1)=1+m$ spacetime split\footnote{In his proof\cite{tsamparlis1979cosmological} Tsamparlis presented a more general split  of the form $m+(n-m)$. However, for our purposes the classical $1+(n-1)$ split will do, since this is relevant to Cosmology.} where $m$ is the number of spatial dimensions, and apply the Cosmological Principle to torsion\footnote{This means to impose a vanishing Lie derivative on the torsion tensor, i.e. $\Lie_{\xi}S_{\mu\nu}^{\;\;\;\lambda}=0$. } then for $m\neq 3$ the only non-vanishing components of torsion read\cite{tsamparlis1979cosmological}
	\beq
	S_{01}^{\;\;\;1}=S_{02}^{\;\;\;2}=S_{03}^{\;\;\;3}=...=S_{0m}^{\;\;\;m} \;\;\; (no\;sum)
	\eeq
	which are essentially described by one degree of freedom. For $m=3$, there also exists the possibility for a completely antisymmetric torsion  part and the only non-vanishing components are
		\beq
	S_{01}^{\;\;\;1}=S_{02}^{\;\;\;2}=S_{03}^{\;\;\;3} 
	\eeq
	\beq
S_{ijk} \propto \epsilon_{ijk}	\label{axto}
	\eeq
	and therefore, in a $4-dimensional$ FLRW Universe, torsion contributes two degrees of freedom. Focusing on non-metricity now, it was shown in \cite{minkevich1998isotropic} that in this case, the non-vanishing components of the latter in a $4-dim$ FLRW Universe are 
	\beq
	Q_{000}
	\eeq
	\beq
	Q_{011}=Q_{022}=Q_{033}
	\eeq
	\beq
	Q_{110}=Q_{220}=Q_{330}
	\eeq
	and we see that the non-metricity is essentially described by three independent degrees of freedom. We should note that each of the five components of torsion and non-metricity can depend only on time as any other coordinate dependence would break isotropy and therefore it is not allowed in such a spacetime. Now, ideally, one would like to have a hypermomentum model that naturally produces the $5$ aforementioned degrees of freedom for torsion and non-metricity in an FLRW Cosmology. This is the main purpose of this paper, and as we will show, the hypermomentum model we propose consists of exactly $5$ degrees of freedom which naturally split into $2$ for torsion and $3$ for non-metricity. In addition, in order to maintain generality we also present the model for arbitrary spacetime dimensions and discuss the differences that arise compared to the special $n=4$ case. Of course writing down the general ansatz for the hypermomentum in an FLRW Universe is, by itself, somewhat trivial. To construct a valid model  we should also equip it with a set of conservation laws that specify the evolution of these non-Riemannian degrees of freedom. Therefore our task would be to first write down the most general form for hypermomentum in a homogeneous Cosmology and then impose the proper conservation laws governing the evolution of it.

	\section{Application of the Unconstrained Hyperfluid Model to Cosmology} 
	Let us see now how the aforementioned unconstrained hyperfluid model fits into a Cosmological setting and also discuss the  modifications and extensions one should consider in order to obtain a Cosmological Hyperfluid that produces all the degrees of freedom of torsion and non-metricity in Cosmology. Recall that in this model the hypermomentum tensor has the form
	\beq
	\Delta_{\alpha\mu\nu}=J_{\alpha\mu}u_{\nu}
	\eeq 
	Let us now consider a homogeneous Cosmological setup. As usual, considering an observer with a normalized  n-velocity field ($u_{\mu}u^{\mu}=-1$) we define the projection operator and temporal  derivative 
	\beq
	h_{\mu\nu}=g_{\mu\nu}+u_{\mu}u_{\nu}
	\eeq
	\beq
\dot{}=u^{\alpha}\nabla_{\alpha}
\eeq
	which constitute an $1+(n-1)$ spacetime split.\footnote{In \cite{iosifidis2018raychaudhuri}, in deriving the generalized Raychaudhuri equation with torsion and non-metricity we considered a more general split when the velocity field was not assumed to be normalized. Then the projection also changes accordingly\cite{iosifidis2018raychaudhuri}.} Then, the most general form of $J_{\alpha\mu}$ in such a setting reads
	\beq
	J_{\alpha\mu}=\phi h_{\alpha\mu}+\omega u_{\alpha}u_{\mu}
	\eeq
	where  $\phi(t)$ and $\omega(t)$ are functions of time only. It is obvious then that in this case the hypermomentum, being of the form
	\beq
	\Delta_{\alpha\mu\nu}=\phi h_{\alpha\mu}u_{\nu}+\omega u_{\alpha}u_{\mu}u_{\nu} \label{Unco}
	\eeq
	can reproduce only two out of the  five (or four) non-Riemannian degrees of freedom. In addition, in this case
	\beq
	\Delta_{[\alpha\mu]\nu}=0  \label{notor}
	\eeq
	and therefore it is difficult to excite torsional degrees of freedom, at least in standard models. Indeed, let us consider the torsional model studied in \cite{kranas2019friedmann}. In this theory we start with the usual Einstein-Hilbert action, assume vanishing non-metricity and a fluid that produces torsion of the form
	\beq
	S_{\alpha\mu\nu}=2u_{[\alpha}h_{\nu]\alpha} \Phi(t) \label{ttt}
	\eeq
	The connection field equations of the Theory read
	\begin{equation}
2\Big( g^{\mu\nu}S_{\lambda}-S^{\mu}\delta^{\nu}_{\lambda}+g^{\mu\sigma}S_{\sigma\lambda}^{\;\;\;\;\nu} \Big)=\kappa \Delta_{\lambda}^{\;\;\mu\nu}
	\end{equation}
	with this we can compute the form of hypermomentum that produces such kind of torsion
	\beq
	\kappa \Delta_{\alpha\mu\nu}=4 \Phi \Big(u_{\alpha}g_{\mu\nu}-u_{\mu}g_{\alpha\nu} \Big) =	\kappa \Delta_{[\alpha\mu]\nu}
	\eeq
	Comparing the latter with ($\ref{notor}$) we see that the unconstrained hyperfluid is incapable of producing the torsion form  $(\ref{ttt})$. It is also evident that ($\ref{Unco}$)	cannot reproduce the pseudoscalar form (\ref{axto}) either. In addition, the hypermomentum $(\ref{Unco})$ in an FLRW Universe is either purely dilatonic (for $\omega=-\phi$) or is disformaly related to matter of dilatonic type. This is readily seen by writing
	\beq
	\Delta_{\alpha\mu\nu}=\hat{g}_{\alpha\mu}\Delta_{\nu}\;\;, \;\;\; \Delta_{\nu}:=\Delta_{\alpha\mu\nu}g^{\alpha\mu}
	\eeq	
	where
	\beq
	\hat{g}_{\mu\nu}=\frac{1}{(n-1)\phi -\omega}\Big[\phi g_{\mu\nu}+(\phi+\omega)u_{\mu}u_{\nu} \Big]
	\eeq
	 is the deformed metric. This kind of matter (dilatonic) produces non-metricity that in a standard Cosmological setting always accelerates Gravitational collapse. We demonstrate this in greater detail in \cite{TBP}. This behaviour had also reported earlier in the work \cite{obukhov1997irreducible}. However, one would expect to have more possibilities and also have contributions coming from shear as well. In this direction, we shall now generalize the form of the hypermomentum so as to encompass not only dilatonic but also torsional and shear contributions.

	\section{Novel Model for Cosmological Perfect Hyperfluid}
	
	To construct our Cosmological Hyperfluid Model, firstly we must restrict the form of the hypermomentum in order to satisfy the Cosmological Principle and secondly we must independently equip the model with a set of conservation laws for the energy-momentum and hypermomentum tensors. Let us start by deriving the allowed form of hypermomentum in an FLRW Universe.

	 Our starting point is to use the high symmetry of the FLRW spacetime and constrain the  form of the hypermomentum in such a space. This can be achieved by demanding a vanishing Lie derivative of all objects living in the host spacetime. As we already mentioned, this procedure was followed in \cite{tsamparlis1979cosmological} in order to obtain the allowed forms of torsion in an FLRW Universe and in \cite{minkevich1998isotropic} for the allowed non-metricity.  Applying this to the hypermomentum tensor we have
	\beq
	\Lie_{\xi}\Delta_{\lambda}^{\;\;\;\mu\nu}=0
	\eeq
	or equivalently
	\beq
	\Lie_{\xi}\Delta_{\alpha\mu\nu}=0
	\eeq
	where we have used the fact that $\Lie_{\xi}g_{\alpha\lambda}=0$.  The above demand implies that\footnote{The proof is direct a generalization of the proof of outlined in \cite{tsamparlis1979cosmological} in order to obtain the possible forms of torsion in Cosmology, the only difference being that we now apply it to a general type-($0,3$) tensor without any symmetry. For more details on this proof (in the case of torsion), the reader is refereed to \cite{tsamparlis1979cosmological}.}
	\beq
	\Delta_{i00}=\Delta_{0i0}=\Delta_{00i}=0
	\eeq
	This can be easily understood, since any of the above terms given that it was not vanishing would create a preferred spatial direction and therefore destroy the isotropy of the model. In addition we have
	\beq
	\Delta_{ijk}=0
	\eeq
	for $m\neq 3 $ and
	\beq
		\Delta_{ijk}\propto \epsilon_{ijk}
	\eeq
	for $m=3$ where $m=n-1$ denotes spatial space dimensionality.
	With the above results and given the fact that the only building blocks of an FRW model are the metric (or the projector $h_{\mu\nu}$) and the four velocity $u^{\mu}$, we state the most general covariant form of the hypermomentum to be
	\beq
	\Delta_{\alpha\mu\nu}=\phi g_{\mu\alpha}u_{\nu}+\chi g_{\nu\alpha}u_{\mu}+\psi u_{\alpha}g_{\mu\nu}+\tilde{\omega} u_{\alpha}u_{\mu} u_{\nu}+\epsilon_{\alpha\mu\nu\kappa}u^{\kappa}\zeta
	\eeq 
	for $m=3$ and
	\beq
\Delta_{\alpha\mu\nu}=\phi g_{\mu\alpha}u_{\nu}+\chi g_{\nu\alpha}u_{\mu}+\psi u_{\alpha}g_{\mu\nu}+\tilde{\omega} u_{\alpha}u_{\mu} u_{\nu}
\eeq 
	for $m\neq 3$,	
	where $\phi,\chi,\psi, \tilde{\omega}$ and $\zeta$ are functions of time. The above can equivalently be expressed as
	\beq
	\Delta_{\alpha\mu\nu}=\phi h_{\mu\alpha}u_{\nu}+\chi h_{\nu\alpha}u_{\mu}+\psi u_{\alpha}h_{\mu\nu}+\omega u_{\alpha}u_{\mu} u_{\nu} +\epsilon_{\alpha\mu\nu\kappa}u^{\kappa}\zeta \;\;,\;\; m=3
	\eeq
		\beq
	\Delta_{\alpha\mu\nu}=\phi h_{\mu\alpha}u_{\nu}+\chi h_{\nu\alpha}u_{\mu}+\psi u_{\alpha}h_{\mu\nu}+\omega u_{\alpha}u_{\mu} u_{\nu} \;\;,\;\; m\neq 3
	\eeq	
	\beq
	 \omega=\tilde{\omega}-\phi -\psi-\chi
	\eeq
	which turns out to be much more convenient for the calculations. We may combine the above into the single expression
		\beq
	\Delta_{\alpha\mu\nu}^{(n)}=\phi h_{\mu\alpha}u_{\nu}+\chi h_{\nu\alpha}u_{\mu}+\psi u_{\alpha}h_{\mu\nu}+\omega u_{\alpha}u_{\mu} u_{\nu}+\delta_{n,4}\epsilon_{\alpha\mu\nu\kappa}u^{\kappa}\zeta \label{Dform}
	\eeq
	where $\delta_{n,4}$ is Kronecker's delta.  This is the most general covariant form of the hypermomentum tensor that is in accordance with the Copernican Principle and thus the most general form of hypermomentum for FRW Universes and (to the best of our knowledge) it is presented here for the first time. Any Cosmological fluid carrying torsion and non-metricity would be a subcase of the latter. Note that the evolution of hypermomentum is governed by $5$ ($4$ for $m\neq3$) functions of time. This is in accordance with the fact that the symmetry of FRW allows only for $2$ ($1$ for $m\neq3$) torsional and $3$ non-metric components. Our result is in perfect agreement with these facts since the above $5$ ($4$  for $m\neq3$) degrees of freedom for hypermomentum  beautifully split into the $2+3$ ($1+3$ for $m\neq3$) components for torsion and non-metricity respectively. This is easily seen by decomposing hypermomentum into its antisymmetric, dilatonic, and proper shear parts according to 
\beq
\Delta_{[\alpha\mu]\nu}=(\psi-\chi)u_{[\alpha}h_{\mu]\nu}+\delta_{n,4}\epsilon_{\alpha\mu\nu\kappa}u^{\kappa}\zeta
\eeq
\beq
\Delta_{\nu}:=\Delta_{\alpha\mu\nu}g^{\alpha\mu}=\Big[ (n-1) \phi -\omega\Big] u_{\nu} \label{dil}
\eeq
\beq
\breve{\Delta}_{\alpha\mu\nu}=\Delta_{(\alpha\mu)\nu}-\frac{1}{n}g_{\alpha\mu}\Delta_{\nu} =\frac{(\phi+\omega)}{n}\Big[ h_{\alpha\mu}+(n-1)u_{\alpha}u_{\mu} \Big] u_{\nu} +(\psi +\chi)u_{(\mu}h_{\alpha)\nu}
\eeq
respectively. The antisymmetric part is known to be related with the spin part of torsion\cite{hehl1995metric} while the dilation and proper shear (symmetric traceless part) are most often  related to non-metricity. From the above decompositions we see that two degrees of freedom  are related to torsion (or one for $m\neq 3$) and the remaining three are reserved for non-metricity. To be more specific, from the above we see that two degrees of freedom (or one dof for $m=3$) are given to the antisymmetric (in the first two indices) and completely antisymmetric part of hypermomentum, another degree of freedom is given to the dilation (trace) and two degrees of freedom govern the evolution of shear (symmetric traceless part of hypermomentum). Of course the above results for the covariant form of the hypermomentum can also be applied to the distortion tensor and one finds
\beq
N_{\alpha\mu\nu}^{(n)}=X(t)u_{\alpha}h_{\mu\nu}+Y(t)u_{\mu}h_{\alpha\nu}+Z(t)u_{\nu}h_{\alpha\mu}+V(t)u_{\alpha}u_{\mu}u_{\nu} +\epsilon_{\alpha\mu\nu\lambda}u^{\lambda}W(t)\delta_{n,4}
\eeq
representing the most general form of distortion tensor in an FLRW Universe.	
Again	W(t) is a pseudoscalar that exists only for $m=3$. In addition, we may express the covariant forms for torsion and non-metricity as
	\beq
Q_{\alpha\mu\nu}=A(t)u_{\alpha}h_{\mu\nu}+B(t) h_{\alpha(\mu}u_{\nu)}+C(t)u_{\alpha}u_{\mu}u_{\nu}  \label{Qnmcos2}
\eeq
\beq
S_{\mu\nu\alpha}^{(n)}=2u_{[\mu}h_{\nu]\alpha}\Phi(t)+\epsilon_{\mu\nu\alpha\rho}u^{\rho}P(t)\delta_{n,4}
\eeq
 Note that the functions determining the distortion  are linearly related with the functions of torsion and non-metricity! This is easily shown by using the relations
\beq
Q_{\nu\alpha\mu}=2 N_{(\alpha\mu)\nu} \;\;, \;\;
S_{\mu\nu\alpha}=N_{\alpha[\mu\nu]}
\eeq	
which result in
\beq
2(X+Y)=B \;, \;\; 2Z=A\;, \;\; 2V=C \;, \;\; 2\Phi =Y-Z\;, \;\; P = W	
\eeq
or inverting them
\beq
W=P \;, \;\; V=C/2 \;, \;\; Z=A/2	
\eeq
\beq
Y=2\Phi +\frac{A}{2}	\;\;, \;\;\; X=\frac{B}{2}- 2 \Phi -\frac{A}{2}
\eeq
Given that the matter Lagrangian is linear in the connection and the Gravitational sector is linear in the curvature and mostly quadratic in $S_{\alpha\mu\nu}$ and $Q_{\mu\nu}$,  the above functions spanning torsion and non-metricity would be algebraically related (and produced by) with the functions that parametrize the hypermomentum.

\subsection{Conservation Laws}

In order to derive the Conservation Laws for our hyperfluid model we will, as usual, employ the diffeomorphism invariance of the matter action. These conservation laws for general MAG Theories have been obtained in \cite{obukhov2013conservation,obukhov2014conservation}\footnote{We also derive a mixing of these conservation laws in the Appendix. Note also that their form when matter decouples from the connection (Palatini) and the subsequent conservation of the energy-momentum tensor was given in \cite{koivisto2006note}. }. In our notations they read
\beq
\frac{1}{\sqrt{-g}}\hat{\nabla}_{\mu}(\sqrt{-g}t^{\mu}_{\;\;\alpha})=\frac{1}{2} \Delta^{\lambda\mu\nu}R_{\lambda\mu\nu\alpha}+\frac{1}{2}Q_{\alpha\mu\nu}T^{\mu\nu}+2 S_{\alpha\mu\nu}t^{\mu\nu}
\eeq
	\beq
t^{\mu}_{\;\;\lambda}= T^{\mu}_{\;\;\lambda}-\frac{1}{2 \sqrt{-g}}\hat{\nabla}_{\nu}(\sqrt{-g}\Delta_{\lambda}^{\;\;\mu\nu}) ,
\;\;\;\;
\hat{\nabla}_{\nu}:=2S_{\nu}-\nabla_{\nu}
\eeq
Note that the second of the above is exactly the relation we obtained in the previous section connecting the three energy related tensors. In addition, as seen from the first one, the metrical energy momentum naturally couples to non-metricity while the canonical couples to torsion. With the help of the above we will construct the conservation laws for our model. Our basic assumption is that the metrical coincides with the canonical energy-momentum tensor, that is
\beq
t^{\mu\nu}=T^{\mu\nu} \label{tT}
\eeq
Note that this was also a consequence of the unconstrained hyperfluid of \cite{obukhov1993hyperfluid}. One could argue that ($\ref{tT}$) could not be so since the CEMT is asymmetric in general while the MEMT is symmetric by construction. However we should recall that our model is tailored for homogeneous Cosmologies, in which any antisymmetric two index object vanishes identically. This means that $t_{[\mu\nu]}\equiv 0$ in a homogeneous Cosmological setup and therefore there is only the symmetric part it $t_{\mu\nu}=t_{(\mu\nu)}$ which we assume to be equal to $T_{\mu\nu}$. Having clarified this, the conservation laws for our model take their final form
\beq
\tilde{\nabla}_{\mu}T^{\mu}_{\;\;\nu}=\frac{1}{2} \Delta^{\alpha\beta\gamma}R_{\alpha\beta\gamma\nu} \label{ConLaw1}
\eeq
\beq
\hat{\nabla}_{\nu}\Big( \sqrt{-g}\Delta_{\lambda}^{\;\;\;\mu\nu}\Big)=0 \label{ConLaw2}
\eeq
where in arriving at the above we have also expanded $\tilde{\nabla}_{\mu}T^{\mu}_{\;\;\nu}$ into its Levi-Civita part plus non-Riemannian contributions.

It is also worth stressing out that the dilatonic part $\Delta^{\nu}$ always scales as $\frac{1}{a^{n-1}}$ regardless of the Gravitational action\footnote{Of course the form of $a(t)$ will depend upon the Gravitational action, but the relation between the dilation and the scale factor will always the same as we prove.}. This is easily proved by contracting the conservation law ($\ref{ConLaw2}$) of hypermomentum in $\mu,\lambda$ to obtain
\beq
\hat{\nabla}_{\nu}(\sqrt{-g}\Delta^{\nu})=0
\eeq
or equivalently
\beq
 \partial_{\nu}(\sqrt{-g}\Delta^{\nu})=0 \label{dilc}
\eeq
that is
\beq
\Delta^{\nu}=\frac{c_{0}u^{\nu}}{a^{n-1}}
\eeq
where $c_{0}$ is some integration constant. In addition, comparing the latter with $(\ref{dil})$ we get the constraint
\beq
(n-1) \phi -\omega=\frac{c_{0}}{a^{n-1}}
\eeq
The above equation is kinematical in the sense that its form is independent of the Gravitational action that we consider, namely it is valid for any Theory. It is also worth pointing out that if the Gravitational part of the action depends explicitly only on the curvature tensor (and the metric of course) then equation $(\ref{dilc})$ holds true identically not only for our hyperfluid model but for any Theory. Indeed, any gravitational action of the form
\beq
S_{G}=S[g_{\mu\nu},R^{\lambda}_{\;\;\alpha\beta\gamma}(\Gamma,\partial \Gamma)]
\eeq
is invariant under special projective transformations, the so-called $\lambda$-transformations
\beq
\Gamma^{\lambda}_{\;\;\mu\nu}\longrightarrow \Gamma^{\lambda}_{\;\;\mu\nu}+\delta^{\lambda}_{\mu}\partial_{\nu}\lambda
\eeq
since, as can be trivially checked, the Riemann tensor remains invariant ($R^{\lambda}_{\;\;\alpha\beta\gamma}\rightarrow$ $R^{\lambda}_{\;\;\alpha\beta\gamma}$ ) under the above connection transformation.
Then implementing this to the matter action, up to surface terms we obtain the constraint
\beq
\partial_{\mu}(\sqrt{-g}\Delta^{\mu})=0
\eeq
which has the form of a conservation law (see also \cite{ponomariov1982generalized,iosifidis2019scale}) and is valid for any model, not just for our Cosmological Hyperfluid.
Let us now try to see the conservation law from another perspective and ask the following. What restrictions does the conservation law ($\ref{ConLaw2}$) imposes on the underlying Geometry and under what circumstances does it hold as an identity without affecting the Geometry? Phrasing it differently we may ask, what kind of invariance could impose the conservation law ($\ref{ConLaw2}$) identically? Giving it a little though we come up with such a transformation. Consider the connection transformation\footnote{ Note also that the class of special projective transformations is included in $(\ref{NabD})$. Indeed, for $\xi^{\lambda}_{\;\;\mu}=\delta^{\lambda}_{\;\;\mu}\lambda$ where $\lambda$	is an arbitrary scalar, the above connection transformations becomes a special projective. In this sense, a Theory that is invariant under ($\ref{NabD}$) is also invariant under special projective transformations. Of course, it is evident that the converse is not true.}
	\beq
\Gamma^{\lambda}_{\;\;\;\mu\nu} \longrightarrow \Gamma^{\lambda}_{\;\;\;\mu\nu}  +\nabla_{\nu}\xi^{\lambda}_{\;\;\mu} \label{NabD}
\eeq
where $\xi^{\lambda}_{\;\;\mu}$ is an arbitrary type $(1,1)$ tensor field. Then, if the Gravitational part is invariant under the above transformation, this invariance is carried over to the matter part and one has
\beq
\delta S_{M}=0=-\frac{1}{2}\int d^{n}x\sqrt{-g}\Delta_{\lambda}^{\;\;\mu\nu} \nabla_{\nu}\xi^{\lambda}_{\;\;\mu}
\eeq
Now, partially integrating, disregarding surface terms and using the fact that $\xi^{\lambda}_{\;\;\mu}$ is arbitrary, the above invariance gives us the desired conservation law
	\beq
\hat{\nabla}_{\nu}\Big( \sqrt{-g}\Delta_{\lambda}^{\;\;\;\mu\nu}\Big)=0 \label{DD1}
\eeq
Note that if the action is not invariant under ($\ref{NabD}$) additional constraints must be imposed on the geometry so as to obtain ($\ref{DD1}$). To see this let us consider a MAG Theory consisting of the usual Einstein Hilbert term and some matter fields that carry hypermomentum. Then, considering the transformation ($\ref{NabD}$) where $\xi^{\lambda}_{\;\;\mu}$ is an infinitesimal, the Riemann tensor changes according to
\beq
R^{\mu}_{\;\;\nu\alpha\beta} \longrightarrow R^{\mu}_{\;\;\nu\alpha\beta}+R^{\mu}_{\;\;\lambda\alpha\beta}\xi^{\lambda}_{\;\;\nu}-R^{\lambda}_{\;\;\nu\alpha\beta}\xi^{\mu}_{\;\;\lambda}
\eeq	
where in arriving at the above we have used the Bianchi identity for $\nabla_{[\alpha}\nabla_{\beta]}\xi^{\lambda}_{\;\;\mu}$ and have kept only linear terms in $\xi^{\lambda}_{\;\;\mu}$. With this we can see that the Ricci scalar transforms as
\beq
R \rightarrow R+(R_{\mu\nu}+\check{R}_{\nu\mu})\xi^{\mu\nu}
\eeq 	
From this it is apparent that for the geometries for which 
\beq
R_{\mu\nu}=-\check{R}_{\nu\mu} \label{RRtil}
\eeq
 the Ricci  scalar remains invariant under ($\ref{NabD}$). When the Gravitational sector respects this invariance this is also applied to the matter sector of the Theory and we get the conservation law $(\ref{DD1})$ for hypermomentum.
Put another way, if the Theory at hand is invariant under ($\ref{NabD}$) then the conservation law ($\ref{DD}$) follows as a consequence of this invariance. If it is not invariant, additional constraints must be imposed on the underlying geometry to make it so. In fact for our above example we can check this statement explicitly. Consider the Theory
\beq
S[g,\Gamma,\phi]=\frac{1}{2\kappa}\int d^{n}x \sqrt{-g} R+ S_{M}[g,\Gamma,\phi]
\eeq
where the first term is the usual Einstein-Hilbert action, $S_{M}$ denotes the matter sector and $\phi$ the matter fields that produce spacetime torsion non-metricity. Variation of the latter with respect to the metric and the independent connection, yields the set of field equations
\beq
R_{(\mu\nu)}-\frac{1}{2}Rg_{\mu\nu}=\kappa T_{\mu\nu}
\eeq
\beq
-\frac{\nabla_{\lambda}(\sqrt{-g}g^{\mu\nu})}{\sqrt{-g}}+\frac{\nabla_{\sigma}(\sqrt{-g}g^{\mu\sigma})\delta^{\nu}_{\lambda}}{\sqrt{-g}} \\
+2(S_{\lambda}g^{\mu\nu}-S^{\mu}\delta_{\lambda}^{\nu}+g^{\mu\sigma}S_{\sigma\lambda}^{\;\;\;\;\nu})=\kappa \Delta_{\lambda}^{\;\;\mu\nu} \label{fe2}
\eeq
After some lengthy but rather straightforward calculations we finally arrive at
\begin{gather}
\hat{\nabla}_{\nu}\Big( -\nabla_{\lambda}(\sqrt{-g}g^{\mu\nu})+\nabla_{\sigma}(\sqrt{-g}g^{\mu\sigma})\delta^{\nu}_{\lambda} 
+\sqrt{-g}(S_{\lambda}g^{\mu\nu}-S^{\mu}\delta_{\lambda}^{\nu}+g^{\mu\sigma}S_{\sigma\lambda}^{\;\;\;\;\nu}) \Big)=-\sqrt{-g}g^{\mu\nu}(\check{R}_{\nu\lambda}+R_{\lambda\nu})
\end{gather}
or employing the connection field equation
\beq
\hat{\nabla}_{\nu}\Big( \sqrt{-g}\Delta_{\lambda}^{\;\;\;\mu\nu}\Big)=-\sqrt{-g}g^{\mu\nu}(\check{R}_{\nu\lambda}+R_{\lambda\nu}) \label{DD2}
\eeq
From which we conclude that for the class of geometries for which $(\ref{RRtil})$ holds true, the law ($\ref{DD1}$) follows. We should note that for an FLRW Universe and for vanishing non-metricity the right hand side of ($\ref{DD2}$) vanishes independently making the left hand side of the latter equation an identity. We will see the implications of this in the sequel. Moreover, it is also worth mentioning that the above restriction on geometries is also imposed on the unconstrained hyperfluid model \cite{obukhov1996model} since the conservation law (\ref{DD1}) we impose is essentially the same and only the covariant form of the hypermomentum differs between the two models. This change however, is very crucial and predicts a totally different dynamics as we explicitly show in what follows.

\subsection{Conservation Laws in FLRW Universes}
Notice that in deriving the conservation laws of the previous section we made no assumption about the symmetries of the space. In order to obtain their form for homogeneous Cosmologies we combine equations ($\ref{Tform}$) and ($\ref{Dform}$) with $({\ref{ConLaw1}})$ and ($\ref{ConLaw2}$) to arrive at  
\beq
\Big[ \dot{\rho}+(n-1)H(\rho+p) \Big] u_{\nu}+(\rho +p)u^{\mu}\tilde{\nabla}_{\mu}u_{\nu}=\frac{1}{2}u^{\mu}(\phi \hat{R}_{\mu\nu}+\chi R_{\mu\nu}+\psi \check{R}_{\mu\nu}) \label{cl1}
\eeq
\begin{gather}
-\delta^{\mu}_{\lambda}    \frac{\partial_{\nu}(\sqrt{-g}\phi u^{\nu})}{\sqrt{-g}}-u^{\mu}u_{\lambda}      \frac{\partial_{\nu}\Big(\sqrt{-g}(\phi+\chi +\psi +\omega) u^{\nu}\Big)}{\sqrt{-g}}
\nonumber \\
+\left[ \Big(2 S_{\lambda}+\frac{Q_{\lambda}}{2}\Big)u^{\mu}-\nabla_{\lambda}u^{\mu} \right]\chi +\left[ \Big(2 S^{\mu}+\frac{Q^{\mu}}{2}-\tilde{Q}^{\mu}\Big)u_{\lambda}-g^{\mu\nu}\nabla_{\nu}u_{\lambda}\right]\psi
\nonumber \\
+ u^{\mu}u_{\lambda}(\dot{\chi}+\dot{\psi}) -(\phi+\chi+\psi+\omega)(\dot{u}^{\mu}u_{\lambda}+u^{\mu}\dot{u}_{\lambda}) 
=0   \label{conl2}
\end{gather}
 The dynamics of the Perfect Cosmological Hyperfluid is contained in the above two equations. We can simplify $(\ref{cl1})$ even further by first noticing that the homothetic curvature (being the field strength of $Q_{\mu}$) identically vanishes for homogeneous Cosmologies $\hat{R}_{\mu\nu}\equiv 0$ and also that $u^{\nu}u^{\alpha}\tilde{\nabla}_{\alpha}u_{\nu}=0$ since the Levi-Civita connection is metric compatible. Then, contracting ($\ref{ConLaw1}$) with $u^{\nu}$ we get
\beq
\dot{\rho}+(n-1)H(\rho+p)=-\frac{1}{2}u^{\mu}u^{\nu}(\chi R_{\mu\nu}+\psi \check{R}_{\mu\nu})\label{rhop}
\eeq
 Some comments are now in order. Firstly, from equation ($\ref{ConLaw2}$) it is clear that the evolution of the non-Riemannian degrees of freedom does not interfere with the perfect fluid elements ($\rho$,p) but on the other hand, the perfect fluid continuity equation receives additional contributions coming from hypermomentum as seen from $(\ref{rhop})$ (see also eq. $(\ref{ConLaw1}))$.  This was not so for the unconstrained hyperfluid model where one can easily see that for the hypermomentum of the form $\Delta_{\alpha\beta\gamma}=J_{\alpha\beta}u_{\gamma}$ the contribution $\Delta^{\alpha\beta\gamma}R_{\alpha\beta\gamma\nu}$ vanishes identically  for FLRW spacetimes and one arrives at the classic continuity equation $\tilde{\nabla}_{\mu}T^{\mu}_{\;\;\nu}=0$. Note that the additional contribution we get on the right-hand side is in agreement (when contacted with $u^{\nu}$) with \cite{kranas2019friedmann} for the purely torsion case. 
Secondly, we see that the pseudoscalar torsional degree of freedom $\zeta$ contains no direct dynamics since it is totally absent from $(\ref{conl2})$.

 Another point worth mentioning is that due to the high symmetry of the FLRW spacetime, the conservation law $(\ref{conl2})$ or equivalently $(\ref{ConLaw2})$ contains only two independent evolution equations for the $5$ (or $4$) fields describing how hypermomentum develops with the passing of time. In order to get a complete description for all degrees of freedom one must provide three equations of state relating the fields $\phi,\chi,\psi,\omega$ and $\zeta$ with one another. Note that this is really no different from the classical  prefect fluid description of GR (without hypermomentum) where the conservation law for the energy momentum tensor $\tilde{\nabla}_{\mu}T^{\mu\nu}=0$ really contains only one equation (the continuity equation) in an FLRW spacetime. Recall that in this case this is the continuity equation which gives the evolution of fluid's density $\rho$ and one must also impose an equation of state of the form $p=w \rho$ in order to close the system of equations and obtain the evolution for the fluid's pressure $p$ as well. Similarly, in the case of the Perfect Cosmological Hyperfluid, one must provide $3$ (or $2$ for $m\neq 3$) equations of state relating the hypermomentum components, which when combined with $(\ref{conl2})$ will give the complete description of the hyperfluid. The exact relation between these fields will depend on the hydrodynamic properties of the fluid under investigation much like the case of the classical perfect fluid. In the appendix we demonstrate how one can obtain such equations of state by considering a concrete example of a scalar field coupled to the connection.  Let us now recap our proposed Cosmological Hyperfluid Model with the following proposition.
 \begin{prop}
 There exists a Perfect Cosmological Hyperfluid for which the canonical and the metrical energy-momentum  tensors coincide, and is therefore subject to the following conservation laws 
 \beq
 \tilde{\nabla}_{\mu}T^{\mu}_{\;\;\nu}=\frac{1}{2} \Delta^{\alpha\beta\gamma}R_{\alpha\beta\gamma\nu} 
 \eeq
 \beq
 \hat{\nabla}_{\nu}\Big( \sqrt{-g}\Delta_{\lambda}^{\;\;\;\mu\nu}\Big)=0 
 \eeq
 for the energy-momentum and hypermomentum currents respectively. The former has the classic perfect fluid form 
 \beq
 T_{\mu\nu}=\rho u_{\mu}u_{\nu}+p h_{\mu\nu}
 \eeq 
 while the latter is given by
 \beq
 \Delta_{\alpha\mu\nu}^{(n)}=\phi h_{\mu\alpha}u_{\nu}+\chi h_{\nu\alpha}u_{\mu}+\psi u_{\alpha}h_{\mu\nu}+\omega u_{\alpha}u_{\mu} u_{\nu}+\delta_{n,4}\epsilon_{\alpha\mu\nu\kappa}u^{\kappa}\zeta 
 \eeq
 which is the most general covariant form for the hypermomentum that is compatible with the Cosmological Principle. The above doublet of conservation laws along with the proper  set of equations of state for the various fields, endowed with the field equations of the Theory under consideration, provide a complete guide for the description of non-Riemannian effects (torsion and non-metricity)  in Cosmology. 
 \end{prop}

\subsection{Fluid Motion}

Let us now discuss the motion of the elements of the hyperfluid. As usual the fluid motion is obtained by the conservation laws. In our case as we have seen previously, it holds that
\beq
\tilde{\nabla}_{\mu}T^{\mu}_{\;\;\nu}=\frac{1}{2} \Delta^{\alpha\beta\gamma}R_{\alpha\beta\gamma\nu} \label{tR}
\eeq
Now, expanding the left hand side and using ($\ref{rhop}$) the latter is brought to the form
\beq
\frac{d^{2}x^{\lambda}}{d\lambda^{2}}+\tilde{\Gamma}^{\lambda}_{\;\;\;\mu\nu}\frac{d x^{\mu}}{d \lambda}\frac{d x^{\nu}}{d \lambda}=\frac{1}{2 (\rho +p)} \Delta^{\alpha\beta\gamma}R_{\alpha\beta\gamma\rho}h^{\rho\lambda}
\eeq
where $\lambda$ is an affine parameter and we have also developed $u^{\mu}\tilde{\nabla}_{\mu}u_{\nu}$ and have raised the free index. Note now that in a Cosmological setting, $\Delta^{\alpha\beta\gamma}R_{\alpha\beta\gamma\rho}$ having one free spacetime index can only be proportional to the four-velocity $u_{\rho}$ which, on the other hand, is orthogonal to the projector $h^{\rho\nu}$. Therefore, the right hand side of the above vanishes leaving us with the ordinary geodesic equation
\beq
\frac{d^{2}x^{\lambda}}{d\lambda^{2}}+\tilde{\Gamma}^{\lambda}_{\;\;\;\mu\nu}\frac{d x^{\mu}}{d \lambda}\frac{d x^{\nu}}{d \lambda}=0
\eeq
We therefore conclude that the elements of the hyperfluid continue to follow the Riemannian geodesics. Of course this result came as a consequence of the high symmetry of the FLRW Universes and we do not expect it to hold true for more general spacetimes, as is apparent from ($\ref{tR}$). We now move on to discuss specific applications of the Model.

\section{Friedmann Equations with Torsion and Non-metricity}
Given that we are in the MAG framework, the most straightforward way to generalize the classic Friedmann equations, is to consider the Einstein Hilbert action along with the presence of a hyperfluid which we shall take to be the Perfect Cosmological Hyperfluid. In the following discussion we shall consider a flat ($K=0$) FLRW Universe with the usual Robertson-Walker line element
\beq
ds^{2}=-dt^{2}+a^{2}(t)\delta_{ij}dx^{i}dx^{j}
\eeq
 Let us start our discussion with the case of vanishing non-metricity.

\subsection{The case of pure Torsion}
If we start with the Einstein-Hilbert action, considering vanishing non-metricity and allowing hyperfluid induced torsion we have the following field equations
\beq
R_{(\mu\nu)}-\frac{1}{2}R g_{\mu\nu}=\kappa T_{\mu\nu}	
\eeq
\beq
P_{\lambda}^{\;\;\;\mu\nu}=
\kappa \Delta _{\lambda}^{\;\;\;\mu\nu}
\eeq
Now, since  non-metricity is zero, the Palatini tensor reads
\begin{equation}
P_{\lambda}^{\;\;\;\mu\nu}=2\Big( g^{\mu\nu}S_{\lambda}-S^{\mu}\delta^{\nu}_{\lambda}+g^{\mu\sigma}S_{\sigma\lambda}^{\;\;\;\;\nu} \Big)
\end{equation}
 In addition, as we have seen in an FLRW Universe one of  the two allowed\footnote{Recall that there is also the pseudoscalar degree of freedom which we shall ignore here for the purposes of this example.} forms for torsion is
\beq
S_{\mu\nu\alpha}=2 \phi u_{[\mu}g_{\nu]\alpha}
\eeq
From the above we find the form of hypermomentum that induces this kind of torsion
\beq
\kappa \Delta_{\alpha\mu\nu}=2(n-2) \phi \Big(u_{\alpha}g_{\mu\nu}-u_{\mu}g_{\alpha\nu} \Big) \label{Dtor}
\eeq
and thus for $n=4$
\beq
\kappa \Delta_{\alpha\mu\nu}=\kappa \Delta_{[\alpha\mu]\nu}=4 \phi \Big(u_{\alpha}g_{\mu\nu}-u_{\mu}g_{\alpha\nu} \Big)\label{Dv}
\eeq
Then, considering the post Riemannian expansions of the Ricci tensor and  scalar, and taking the $00$ and $ij$ components of the metric field equations we easily obtain the modified Friedmann equations in the presence of torsion (see also \cite{kranas2019friedmann})
\beq
H^{2}=\frac{\kappa}{3}\rho +4 H \phi  -4\phi^{2}
\eeq
\beq
\frac{\ddot{a}}{a}=-\frac{\kappa}{6}(\rho +3 p) +2 \dot{\phi} +2 H \phi
\eeq
Let us now recall our conservation laws for the perfect Cosmological hyperfluid. They read
\beq
\tilde{\nabla}_{\mu}T^{\mu}_{\;\;\nu}=\frac{1}{2}\Delta^{\alpha\beta\gamma}R_{\alpha\beta\gamma\nu} \label{con1}
\eeq
\beq
\hat{\nabla}_{\nu}\Big( \sqrt{-g}\Delta_{\lambda}^{\;\;\;\mu\nu}\Big)=0 \label{DD}
\eeq
where, as usual
\beq
T_{\mu\nu}=\rho u_{\mu}u_{\nu}+p h_{\mu\nu}
\eeq
and for the given case $\Delta_{\mu\nu\alpha}$ is given by $(\ref{Dtor})$. Now, we readily compute
\beq
\Delta^{\alpha\beta\gamma}R_{\alpha\beta\gamma\nu}=4 \phi \Big[ (n-3)\rho+(n-1)p \Big]u_{\nu}
\eeq
Then,  expanding $(\ref{con1})$ and contracting with $u^{\nu}$ we finally obtain
\beq
\dot{\rho}+3 H(\rho+ p)=2 \phi (\rho +3 p)
\eeq
which is in perfect agreement with the corresponding continuity equation used in \cite{kranas2019friedmann}\footnote{Note the different definitions used there. In particular our $\phi's$ are related through $\phi \rightarrow -\phi$.}. This justifies the self-consistency of our proposed model. Note that the aforementioned conservation law for the density was derived in \cite{kranas2019friedmann} by using the Bianchi identities. Here we obtained it by employing the diffeomorphism invariance of the matter action. The two procedures yield, as expected, the same result with the only difference being that the second method leads us straightforwardly to the conservation law. As far as the evolution of $\phi$ is concerned in this case the conservation law for the hypermomentum trivializes in an FLRW background because the latter is antisymmetric in its first two indices as is clear from ($\ref{Dv}$). This feature of was already pointed out in \cite{tsamparlis1981methods} but no explanation was given there. Here we have already explained why this is so. The reason behind the arbitrariness of $\phi$  traces back to equation ($\ref{DD2}$). As we have mentioned the right-hand side of that equation vanishes identically in an FLRW spacetime if the non-metricity is zero and therefore the left-hand side of the same equation vanishes identically as well without imposing any constraint on the sources. We shall now switch off torsion and focus on the effect of non-metricity.

\subsection{The case of pure Non-metricity}
 Turning now  our attention to  non-metricity, we shall start with the simplest case which is the Weyl non-metricity and then derive the equations for the general case. Considering again a Theory consisting of the Einstein-Hilbert action and a non-metric hyperfluid, the field equations read
 \beq
 R_{(\mu\nu)}-\frac{1}{2}Rg_{\mu\nu}=\kappa T_{\mu\nu}
 \eeq
 \beq
 P_{\lambda}^{\;\;(\mu\nu)}=\kappa \Delta_{\lambda}^{\;\;(\mu\nu)} \label{fe2}
 \eeq
 where $P_{\lambda}^{\;\;(\mu\nu)}$ is the torsion-free symmetrized Palatini tensor, given by
 \beq
 P_{\lambda}^{\;\;(\mu\nu)}=\frac{1}{2}Q_{\lambda}g^{\mu\nu}-Q_{\lambda}^{\;\;\mu\nu}+\left( \tilde{Q}^{(\mu}-\frac{1}{2}Q^{(\mu} \right) \delta^{\nu)}_{\lambda}
 \eeq 
Note the symmetrization in $\mu,\nu$ that occurs now due to the fact that we started with a symmetric connection right from the beginning (i.e. we imposed vanishing torsion).
 \subsubsection{Weyl Non-metricity}
We now assume that the non-metric hyperfluid is such that it produces a Weyl non-metricity of the usual type
\beq
Q_{\alpha\mu\nu}=\frac{1}{n}Q_{\alpha}g_{\mu\nu}=A(t) u_{\alpha}g_{\mu\nu}
\eeq
where $A(t)$ is the function that monitors the effect of non-metricity. This follows as a special case of our Perfect Cosmological Hyperfluid, for which the symmetrized hypermomentum takes the form\footnote{A cautionary remark is in order now. It should be strongly emphasized that one must first make an ansatz for the hypermomentum and subsequently derive the desired form for non-metricity (or torsion) and not the other way around since hypermomentum is the source. Considering an ansatz for non-metricity (or torsion) and then computing the hypermomentum, one runs the risk of loosing degrees of freedom that are originally contained in hypermomentum and subsequently arrive at inconsistent conservation laws.}
\beq
\kappa \Delta_{\lambda}^{\;\;\mu\nu}=\frac{(n-2)}{2 n}\Big[ Q_{\lambda}g^{\mu\nu}-\delta^{(\mu}_{\lambda}Q^{\mu)}\Big]
\eeq
or
\beq
\kappa \Delta_{\alpha\mu\nu}=\frac{(n-2)}{2}A(t)\Big[ u_{\alpha}g_{\mu\nu}-g_{\alpha(\mu}u_{\nu)} \Big]
\eeq
 Following an identical procedure to the one we outlined previously, after some calculations we finally find the modified Friedmann equations in the presence of Weyl non-metricity
\beq
H^{2}=\frac{\kappa}{3}\rho +HA-\frac{A^{2}}{4}
\eeq
\beq
\frac{\ddot{a}}{a}=-\frac{\kappa}{6}(\rho +3 p)+\frac{\dot{A}}{2}+\frac{HA}{2}
\eeq
Note  that the  torsional Friedmann equations of the previous section are mapped to the ones here and vice versa upon the duality exchange
\beq
A \leftrightarrow  4 \phi
\eeq
which holds between vectorial torsion and Weyl non-metricity as we have reported in our previous works\cite{iosifidis2019torsion,iosifidis2018raychaudhuri}(see also \cite{klemm2020einstein}). Now, the conservation laws for the above case read
\beq
\dot{\rho}+3 H(\rho+ p)=\frac{A}{2} (\rho +3 p)
\eeq
while the conservation law for hypermomentum trivializes. This again has to do with the restricted geometry we imposed in this simple example. Indeed, as can be straightforwardly checked, for vanishing torsion and for Weyl non-metricity it holds that $\check{R}_{\mu\nu}+R_{\nu\mu}=0$ in an FLRW spacetime. The latter condition results in the trivialization of the conservation law for hypermomentum  just like the pure torsion case we discussed earlier.  We should note that such trivializations occur only because we have greatly restricted the underlying geometry. Allowing for a richer geometrical structure we get a complete set of equations as we show with the following example.  

\subsubsection{General Non-metricity}
 Let us now allow all three degrees of freedom for non-metricity and present the most straightforward generalization of Friedmann equations in the presence of  non-metricity\footnote{To be more accurate, in the presence of a hyperfluid induced non-metricity.} In the most general case where the non-metricity tensor is given by ($\ref{Qnmcos2}$) after some laborious calculations, from the metric field equations  we finally extract\cite{TBP}
 \beq
 H^{2}=\frac{\kappa}{3}\rho -\frac{1}{2}H\left( \frac{3}{2}B-A+C \right)-\frac{1}{4}\dot{B}+\frac{1}{8}B(A-C)+\frac{1}{4}AC
 \eeq
 \beq
 \frac{\ddot{a}}{a}=-\frac{\kappa}{6}(\rho +3 p)+H\left( A+\frac{C}{2}\right) +\frac{1}{2}\dot{A}-\frac{1}{4}A(A+C)
 \eeq
 These above constitute the most straightforward (minimal) modification of Friedmann equations in the presence of non-metricity.
 More details on the derivation and also on the analysis of the latter we will present elsewhere\cite{TBP}. The above equations are supplemented with the conservation laws ($\ref{ConLaw1}$) and ($\ref{ConLaw2}$). Taking the various components of the aforementioned laws, we find the evolution of $\rho$, $p$ and the hypermomentum. Then, the functions spanning the hypermomentum are algebraically related to the three non-metricity functions as can be easily shown by using the relation between the distortion and hypermomentum\cite{iosifidis2019exactly}. Again, the complete study of this goes beyond the purposes of this paper and will be studied separately\cite{TBP}. As a note we should point out that as can be easily seen from the second Friedmann equation above, the non-metric degrees of freedom could account for accelerated expansion. To clearly see this let us consider the simple special case for which $\dot{A}>0$ and $C=-2A$, then the acceleration equation becomes
 \beq
  \frac{\ddot{a}}{a}=-\frac{\kappa}{6}(\rho +3 p)+\frac{1}{2}\dot{A}+\frac{1}{4}A^{2}
  \eeq
From which it is evident that this kind of non-metricity accelerates the Cosmological expansion. Of course in the general case where all three functions of non-metricity are present, the effect on acceleration equation is not quite totally clear and deserves further investigation.
  We shall now discuss the generalized case where both torsion and non-metricity are allowed.
 
 \section{Friedmann Equations With both Torsion and Non-Metricity}
 Starting again with the Einstein-Hilbert action and the presence of the Cosmological Hyperfluid in the matter sector, leaving the connection general, varying with respect to the metric and the connection, we obtain the field equations
\begin{equation}
R_{(\mu\nu)}-\frac{g_{\mu\nu}}{2}R=\kappa T_{\mu\nu}
\end{equation}
\begin{equation}
P_{\lambda}^{\;\;\;\mu\nu}=\kappa \Delta_{\lambda}^{\;\;\;\mu\nu}
\end{equation}
where
\beq
P_{\lambda}^{\;\;\;\mu\nu}=-\frac{\nabla_{\lambda}(\sqrt{-g}g^{\mu\nu})}{\sqrt{-g}}+\frac{\nabla_{\sigma}(\sqrt{-g}g^{\mu\sigma})\delta^{\nu}_{\lambda}}{\sqrt{-g}} \\
+2(S_{\lambda}g^{\mu\nu}-S^{\mu}\delta_{\lambda}^{\nu}+g^{\mu\sigma}S_{\sigma\lambda}^{\;\;\;\;\nu})
\eeq
is the full Palatini tensor.
After some rather heavy calculations and enough patience we finally find the most general form of the  Friedmann equations with both torsion and non-metricity induced by the presence of the Perfect Cosmological Hyperfluid\cite{TBP2}
\begin{gather}
H^{2}=-\frac{2}{(n-2)}H\left[ \frac{(n-1)}{2}X-\frac{(n-3)}{2}Y+Z+V \right] -\frac{1}{(n-2)}(\dot{X}+\dot{Y})-\frac{1}{(n-2)}(X-Y)(Z+V)+XY \nonumber \\
+\frac{2}{(n-2)}W^{2}\delta_{n,4}+\frac{2 \kappa}{(n-1)(n-2)}\rho
\end{gather}
\beq
\frac{\ddot{a}}{a}=-\frac{\kappa}{(n-1)(n-2)}\Big[ (n-3)\rho +(n-1)p \Big]+\dot{Y}+H(Y+Z+V)-Y(V+Z)
\eeq
where we have expressed the additional contributions in terms of the distortion functions due to computational convenience. Of course these can be written entirely in terms of torsion and non-metricity functions by using the conversion relations of the previous section 
\beq
W=P \;, \;\; V=C/2 \;, \;\; Z=A/2	
\eeq
\beq
Y=2\Phi +\frac{A}{2}	\;\;, \;\;\;
X=\frac{B}{2}- 2 \Phi -\frac{A}{2}
\eeq 
which in turn can be expressed solely in terms of the hypermomentum functions by using the relation between distortion and hypermomentum\cite{iosifidis2019exactly}. The exact derivation of the above generalized Friedmann equations, with all immediate steps,  along with solutions will be presented elsewhere\cite{TBP2}. The generalized Friedmann equations are supplemented with the conservation laws of the Perfect Cosmological Hyperfluid  
\beq
\dot{\rho}+(n-1)H(\rho+p)=-\frac{1}{2}u^{\mu}u^{\nu}(\chi R_{\mu\nu}+\psi \check{R}_{\mu\nu})\label{rhop}
\eeq 
\begin{gather}
-\delta^{\mu}_{\lambda}    \frac{\partial_{\nu}(\sqrt{-g}\phi u^{\nu})}{\sqrt{-g}}-u^{\mu}u_{\lambda}      \frac{\partial_{\nu}\Big(\sqrt{-g}(\phi+\chi +\psi +\omega) u^{\nu}\Big)}{\sqrt{-g}}
\nonumber \\
+\left[ \Big(2 S_{\lambda}+\frac{Q_{\lambda}}{2}\Big)u^{\mu}-\nabla_{\lambda}u^{\mu} \right]\chi +\left[ \Big(2 S^{\mu}+\frac{Q^{\mu}}{2}-\tilde{Q}^{\mu}\Big)u_{\lambda}-g^{\mu\nu}\nabla_{\nu}u_{\lambda}\right]\psi
\nonumber \\
+ u^{\mu}u_{\lambda}(\dot{\chi}+\dot{\psi}) -(\phi+\chi+\psi+\omega)(\dot{u}^{\mu}u_{\lambda}+u^{\mu}\dot{u}_{\lambda}) 
=0 \label{conl22}
\end{gather}

where the former is the modified continuity equation of the perfect fluid components and the latter expresses conservation of the hypermomentum sector of the hyperfluid. We should mention that there is a subtle difference between this case and the cases of vanishing torsion or non-metricity we discussed earlier. When one starts  with a general (unconstrained) affine connection, as it is well known, the Ricci scalar enjoys the projective symmetry
\beq
R \longrightarrow R
\eeq
under projective transformations of the connection
\beq
\Gamma^{\lambda}_{\;\;\mu\nu}\longrightarrow \Gamma^{\lambda}_{\;\;\mu\nu}+\delta^{\lambda}_{\mu}\xi_{\nu}
\eeq
where $\xi_{\nu}$ is an arbitrary one form. The above invariance, when transferred to the matter sector, demands a vanishing dilation current, namely
\beq
\Delta^{\nu}=\Delta_{\lambda}^{\;\;\;\lambda\nu}=\Big[ (n-1)\phi-\omega \Big] u^{\nu}=0
\eeq
and therefore restricts the form of the hyperfluid. However, it should be noted that constraints like the latter have also a pleasing consequence since they provide 'equations of state' between the hypermomentum components\footnote{In this case we get the equation of state $\omega=(n-1)\phi$ relating two of the hypermomentum variables.} which, as mentioned, are appropriate in order to determine the complete Cosmological evolution. Anyway, the case of projective invariance and its relevance to physics  is an open  issue with some recent studies indicating that keeping the invariance is needed in order to avoid ghosts\cite{jimenez2019ghosts,aoki2019scalar} (see also \cite{percacci2019new}). In our case, we see that the projective invariance supplies the model with a desired equation of state among the hyperfluid variables. Below we expand a  little more on the role of invariances and the constraints they put on hypermomentum.

\section{Hypermomentum Matter Types}
As we noted above, a Gravitational action that is invariant under projective transformations will only allow for matter fields with vanishing dilation current (i.e. $\Delta^{\nu}=0$). In fact, one can consider a  more general class of unconstrained vectorial transformations of the form 
\beq
\Gamma^{\lambda}_{\;\;\mu\nu}\longrightarrow \Gamma^{\lambda}_{\;\;\mu\nu}+\delta^{\lambda}_{\mu}\xi_{\nu}+\delta^{\lambda}_{\nu}\zeta_{\mu}+\chi^{\lambda}g_{\mu\nu} \label{vecgen}
\eeq
where $\xi_{\mu}$, $\zeta_{\mu}$, and $\chi_{\mu}$ are in general independent one forms. Some subclasses of the above transformations include
\beq
\Gamma^{\lambda}_{\;\;\mu\nu}\longrightarrow \Gamma^{\lambda}_{\;\;\mu\nu}+\lambda_{1}\delta^{\lambda}_{\mu}v_{\nu}+\lambda_{2}\delta^{\lambda}_{\nu}v_{\mu}+\lambda_{3}v^{\lambda}g_{\mu\nu} \label{veccons}
\eeq
\beq
\Gamma^{\lambda}_{\;\;\mu\nu}\longrightarrow \Gamma^{\lambda}_{\;\;\mu\nu}+\delta^{\lambda}_{\mu}\xi_{\nu} \label{d1}
\eeq
\beq
\Gamma^{\lambda}_{\;\;\mu\nu}\longrightarrow \Gamma^{\lambda}_{\;\;\mu\nu}+\delta^{\lambda}_{\nu}\zeta_{\mu} \label{d2}
\eeq
\beq
\Gamma^{\lambda}_{\;\;\mu\nu}\longrightarrow \Gamma^{\lambda}_{\;\;\mu\nu}+\chi^{\lambda}g_{\mu\nu} \label{d3}
\eeq
where the former is called a constrained vectorial transformation and the latter three
are transformations of the \textit{$1^{st}$, $2^{nd}$ and $3^{rd}$ $kind$} respectively\cite{iosifidis2020linear}.
 In \cite{iosifidis2020linear} we obtained classes of actions that are invariant under the above connection transformations and also derived the constraints imposed on the hypermomentum by such invariances. Combining the above connection transformations with our Cosmological model here we can attribute a clear physical meaning to transformations of the form $(\ref{vecgen})$. In a Cosmological setting, if the Gravitational action is invariant under transformations of the form ($\ref{vecgen}$) (or a subset thereof) then this invariance provides equations of state between the various components of the hyperfluid. To be more specific, by using the results of \cite{iosifidis2020linear} we can state specific equations of state or conservation laws for our Hyperfluid model. We start by reviewing the three propositions we obtained in the aforementioned study and then apply them to the Cosmological Hyperfluid model presented here. 
Consider the Theory 
\beq
S[g,\Gamma,\phi]=\frac{1}{2 \kappa}S_{G}[g,\Gamma]+S_{M}[g,\Gamma,\phi] \label{SG}
\eeq
where $S_{G}$ denotes the Gravitational sector (Geometry) and $S_{M}$ represents the matter fields.
 Set
\beq
\Delta_{(1)}^{\nu}:=\Delta_{\lambda}^{\;\;\lambda\nu}\;, \; \Delta_{(2)}^{\mu}:=\Delta_{\lambda}^{\;\;\mu\lambda}\;, \; \Delta_{(3)\lambda}:=\Delta_{\lambda}^{\;\;\mu\nu}g_{\mu\nu}
\eeq
Then, we get the following three immediate propositions\cite{iosifidis2020linear}

\begin{prop}
	If the gravity sector of ($\ref{SG}$) is invariant under general vectorial transformations of the connection of the form ($\ref{vecgen}$), then the hypermomentum tensor has all of its contractions vanishing, namely
	\beq
	\Delta_{(i)}^{\mu}=0\;, \forall i=1,2,3
	\eeq
	
\end{prop}

\begin{prop}
	If the gravity sector of ($\ref{SG}$) is invariant under constrained vectorial transformations of the form ($\ref{veccons}$), then the traces of the hypermomentum of the Theory satisfy the constraint
	\beq
	\sum_{i=1}^{3}\lambda_{i}\Delta_{(i)}^{\mu}=0
	\eeq
	That is, the three traces of the hypermomentum are  linearly dependent.
\end{prop}

\begin{prop}
	If the gravity sector of ($\ref{SG}$) is invariant under connection transformations of the $i-th$ $kind$ ($i=1,2,3$) then this invariance demands a vanishing $i-th$ contraction of the hypermomentum
	\beq
	\Delta_{(i)}^{\mu}=0
	\eeq
\end{prop}

In the case of the Perfect Cosmological Hyperfluid that we present here, we have
\beq
\Delta_{(1)}^{\nu}:=\Delta_{\lambda}^{\;\;\lambda\nu}=\Big[ (n-1)\phi-\omega \Big] u^{\nu} 
\eeq 
\beq
\Delta_{(2)}^{\mu}:=\Delta_{\lambda}^{\;\;\mu\lambda}=\Big[ (n-1)\chi-\omega \Big] u^{\mu} 
\eeq
\beq
 \Delta_{(3)\lambda}:=\Delta_{\lambda}^{\;\;\mu\nu}g_{\mu\nu}=\Big[ (n-1)\psi-\omega \Big] u_{\lambda}
\eeq
Combining the above results we may  classify some special cases of the Perfect Cosmological Hyperfluid. For instance if the Gravitational sector of a given MAG Theory is invariant under general vectorial transformations of the connection ($\ref{vecgen}$), than the associated Cosmological Hyperfluid satisfies the three 'equations of state'
\beq
\phi=\chi=\psi=\frac{\omega}{(n-1)}
\eeq
We shall call such a fluid, a  completely traceless hyperfluid.   On the other hand, if the Gravitational action is invariant under only one of $(\ref{d1})$,$(\ref{d2})$ or $(\ref{d3})$ then one gets only one hypermomentum equation of state for each case as seen from the above. It is worth mentioning that if the connection transformations are of the special kind, that is the one forms appearing in $(\ref{veccons})$ are all exact, then the corresponding constraints for the hypermomentum turn into conservation laws among its constituents. For instance, a Theory that is invariant under the action of special linear transformations of the form 
\beq
\Gamma^{\lambda}_{\;\;\mu\nu}\longrightarrow \Gamma^{\lambda}_{\;\;\mu\nu}+\delta^{\lambda}_{\mu}\partial_{\nu}X+\delta^{\lambda}_{\nu}\partial_{\mu}Y+g_{\mu\nu} \partial^{\lambda}Z\label{specli}
\eeq
where $X,Y$ and $Z$ are arbitrary independent scalars,
will have an  associated Cosmological Hyperfluid obeying the conservation laws
\beq
\partial_{\mu}\Big[ \sqrt{-g}\Big(  (n-1)\phi-\omega\Big) u^{\mu} \Big]=0\;\;, \;\; \partial_{\mu}\Big[ \sqrt{-g}\Big(  (n-1)\chi-\omega\Big) u^{\mu} \Big]=0\;\;, \;\; \partial_{\mu}\Big[ \sqrt{-g}\Big(  (n-1)\psi-\omega\Big) u^{\mu} \Big]=0
\eeq 
Of course the list could go on and on by considering different types of transformations for the connection and deriving the corresponding constraints on the matter. Our intention here was to illustrate the procedure that one has to follow in order to obtain the constraints of the matter sector. The aforementioned types of matter along with their extensions certainly deserve further investigation so as to unveil their true physical content in Cosmological situations. As a final note let us mention that another interesting case is that of a frame rescaling invariant fluid. In this case the density and pressure of the fluid are related with the hypermomentum variables through (recall eq. ($\ref{tTT}$))
\beq
2(p-3 \rho)=3 H F+\dot{F}\;\;, \;\;\; F:=3\phi -\omega
\eeq
We should mention that this type of fluid is not contained in our Cosmological Hyperfluid desrciption since its canonical and metrical energy-momentum tensors do not coincide. However, it would also be very interesting to study the characteristics of such a fluid especially with regards to its Cosmological effects. All the above aspects certainly deserve further and careful investigation.

\section{Conclusions}
 We have developed a novel model for a Cosmological Hyperfluid, namely a fluid that carries hypermomentum and subsequently induces torsional and non-metric degrees of freedom. In constructing the model we essentially imposed the same conservation laws with the unconstrained hyperfluid\cite{obukhov1996model} but we have considered a different ansatz for the covariant form of the hypermomentum that is compatible with the Cosmological Principle. Therefore even though the  conservation laws have the same covariant form, they predict different dynamics owed to the different form of the hypermomentum. In our approach the hyperfluid is described by the usual energy-momentum tensor of a perfect fluid  which is characterized by the energy density $\rho$ and pressure $p$ of the fluid's elements, as well as by a hypermomentum tensor containing the microscopic (can include macroscopic as well) properties of the matter. The latter is spanned by $5$ or $4$ degrees of freedom depending on the dimensionality of the spatial space.
  Given the ansatz for the Robertson-Walker metric, the Cosmological evolution of the Perfect Cosmological Hyperfluid in $4-dim$ is described by $1+2+5=8$ time functions corresponding to the scale factor ($\textbf{1}$) and the degrees of freedom coming from the  energy-momentum ($\textbf{2}$), and the hypermomentum ($\textbf{5}$). Therefore, the complete Cosmological description in the presence of torsion and non-metricity is achieved by finding the form of  the $8$ functions $a(t),\rho(t),p(t), \phi(t),...$ etc.
  
  As an immediate application of our model we considered a Theory consisting of the usual Einstein-Hilbert action and the presence of our Perfect Cosmological Hyperfluid. Consequently we derived the (most straightforward) generalized Friedmann equations in the presence of torsion and non-metricity. The generalized Friedmann equations we obtained, supplemented with the conservation laws ($\ref{ConLaw1}$) and $(\ref{ConLaw2})$, provide then a complete and minimal tool to study the impact of torsion and non-metricity in Cosmology.
We should note that it is possible to generalize our model in a straightforward way. Indeed, the two assumptions for the Perfect Cosmological Hyperfluid are $1)$ That the Cosmological Principle holds true, and $2) $ That the canonical and metrical energy momentum tensors coincide. Given that we would like to stay in a Cosmological context, we should keep the former assumption but nothing prevents us to relax the latter. Then, our model is readily generalized by relaxing the assumption that the CEMT and MEMT coincide and leave them independent a priori or relate them in a different way. In the last section we discussed the example of the frame rescaling invariant fluid which falls into that category.
It would also be interesting to explore the consequences of such a generalized models.

 As closing remark let us mention that the establishment of the Perfect Cosmological Hyperfluid model opens up a whole plethora of possibilities to  study  non-Riemannian effects in Cosmology. The presence of such a fluid is expected to change the  evolution of the Universe at early stages and therefore provide alternative well motivated inflationary scenarios. In addition for non-standard Gravitational actions (going beyond the Einstein-Hilbert) the Cosmological Hyperfluid could just as well be used in order to formulate late-time Cosmological effects. Metric-Affine $f(R)$ Theories in the presence of Perfect Cosmological Hyperfluids would be typical  representatives of such possibilities. The above points will be addressed in future works.

\section{Acknowledgments}
I would like to thank very much Prof. Yuri N. Obukhov for extremely useful explanations and constructive email discussions on the concept of the hyperfluid. I would also like to thank Tomi S. Koivisto, Christos Tsagas, Anastasios Petkou and Jose Beltran Jimenez for useful discussions and comments.
This research is co-financed by Greece and the European Union (European Social Fund- ESF) through the
Operational Programme «Human Resources Development, Education and Lifelong Learning» in the context
of the project “Reinforcement of Postdoctoral Researchers - 2
nd Cycle” (MIS-5033021), implemented by the
State Scholarships Foundation (IKY).

\appendix

	\section{Diffeomorphisms}
Let us consider the (active) diffeomorphism
\beq
x'^{ \mu}=x^{\mu}+\xi^{\mu}(x) \label{diff}
\eeq
which corresponds to a general(active) infinitesimal coordinate transformation. From the transformation law of a scalar
\beq
\phi'(x')=\phi(x)
\eeq
by expanding 	$x'$ about $x$ we find
\beq
\phi'(x')=	\phi'(x+\xi) \approx \phi'(x)+\xi^{\mu}\partial_{\mu}\phi
\eeq
Then by definition, the scalar field changes according to\footnote{The symbol $\delta$ stands here for functional variations, that is measures the difference of the field $\phi$ computed at two different points but in the same coordinate system $\{x\}$, see also \cite{ortin2004gravity}.}
\beq
\delta\phi:= \phi'(x)-\phi(x)=-\xi^{\mu}\partial_{\mu}\phi
\eeq
Considering now the transformation law of the connection under diffeomorphisms
\beq
\Gamma^{' \lambda}_{\;\;\;\;\mu\nu}=\frac{\partial x^{' \lambda}}{\partial x^{\alpha}}\frac{\partial x^{\beta}}{\partial x ^{' \mu}}\frac{\partial x^{\gamma}}{\partial x^{' \nu}}\Gamma^{' \alpha}_{\;\;\;\beta\gamma}+\frac{\partial x^{' \lambda}}{\partial x^{\alpha}}\frac{\partial^{2}x^{\alpha}}{\partial x^{'\mu}\partial x^{'\nu}}
\eeq
and expanding in the same manner as before, we finally arrive at
\beq
\delta_{\xi}\Gamma^{\lambda}_{\;\;\;\mu\nu}=-\xi^{\alpha}\partial_{\alpha}\Gamma^{\lambda}_{\;\;\;\mu\nu}-\Gamma^{\lambda}_{\;\;\;\alpha\nu}\partial_{\mu}\xi^{\alpha}-\Gamma^{\lambda}_{\;\;\;\mu\alpha}\partial_{\nu}\xi^{\alpha}+\Gamma^{\alpha}_{\;\;\;\mu\nu}\partial_{\alpha}\xi^{\lambda}-\partial_{\mu}\partial_{\nu}\xi^{\lambda}
\eeq
Now, notice that this is in fact a tensor. To see this, we use the definitions, of covariant derivative, torsion and curvature to re-express the above in the equivalent form\footnote{The proof is trivial but somewhat lengthy. It is a matter of sticking to the definitions and rearranging the various terms.}
\beq
\delta_{\xi}\Gamma^{\lambda}_{\;\;\;\mu\nu}=\xi^{\alpha}R^{\lambda}_{\;\;\mu\nu\alpha}-\nabla_{\nu}\nabla_{\mu}\xi^{\lambda}-2\nabla_{\nu}(S_{\mu\alpha}^{\;\;\;\;\lambda}\xi^{\alpha}) \label{diffgamma}
\eeq
which is manifestly covariant and therefore reveals the tensorial character of $\delta_{\xi}\Gamma^{\lambda}_{\;\;\;\mu\nu}$. Similarly, one can show that the under diffeomorphisms ($\ref{diff}$) the metric tensor transforms as
\beq
\delta_{\xi}g_{\mu\nu}=-\xi^{\alpha}\partial_{\alpha}g_{\mu\nu}-2g_{\alpha(\mu}\partial_{\nu)}\xi^{\alpha}
\eeq
or equivalently 
\beq
\delta_{\xi}g_{\mu\nu}=-\xi^{\alpha}\partial_{\alpha}g_{\mu\nu}-\partial_{\nu}\xi_{\mu}+\xi^{\alpha}\partial_{\nu}g_{\mu\alpha}-\partial_{\mu}\xi_{\nu}+\xi^{\alpha}\partial_{\mu}g_{\alpha\nu}
\eeq
Furthermore, using the definition of the covariant derivative, the above can be expressed as
\beq
\delta_{\xi}g_{\mu\nu}=-2\tilde{\nabla}_{(\mu}\xi_{\nu)} \label{diffg}
\eeq
or
\beq
\delta_{\xi}g_{\mu\nu}=-2\nabla_{(\mu}\xi_{\nu)}-2N^{\lambda}_{\;\;\;(\mu\nu)}\xi_{\lambda}
\eeq
where $\nabla_{\mu}$ and $\tilde{\nabla_{\mu}}$ represent the covariant derivative of the general affine connection and the Levi-Civita connection respectively and $N^{\lambda}_{\;\;\;\mu\nu}=\Gamma^{\lambda}_{\;\;\;\mu\nu}-\tilde{\Gamma}^{\lambda}_{\;\;\;\mu\nu}$ is the distortion. It will be more useful to use the expression with the Levi-Civita connection
Let us consider now a matter action
\beq
S_{M}[g,\Gamma,\phi]=\int d^{n}x \sqrt{-g}\mathcal{L}_{M}
\eeq
Then, considering a diffeomorphism, we have
\beq
\delta_{\xi}S_{M}=\int d^{n}x \Big[ \frac{\delta (\sqrt{-g}\mathcal{L}_{M})}{\delta g_{\mu\nu}}\delta_{\xi}g_{\mu\nu}+\frac{\delta (\sqrt{-g}\mathcal{L}_{M})}{\delta \Gamma^{\lambda}_{\;\;\;\mu\nu}}\delta_{\xi}\Gamma^{\lambda}_{\;\;\;\mu\nu}+\frac{\delta (\sqrt{-g}\mathcal{L}_{M})}{\delta \phi}\delta_{\xi}\phi \Big]
\eeq
Now, given that the matter fields $(\phi)$ satisfy their field equations, the last term on the right hand side of the above vanishes. Then, using $(\ref{diffgamma}),(\ref{diffg})$ after some partial integrations and assuming that $\xi^{\alpha}$ and its derivatives vanish at the boundary, we arrive at
\begin{gather}
\delta_{\xi}S_{M}=\int d^{n}x \xi^{\alpha}\Big[ \sqrt{-g}(2 \tilde{\nabla}_{\mu}T^{\mu}_{\;\;\alpha}-\Delta^{\lambda\mu\nu}R_{\lambda\mu\nu\alpha})+\hat{\nabla}_{\mu}\hat{\nabla}_{\nu}(\sqrt{-g}\Delta_{\alpha}^{\;\;\mu\nu})+2S_{\mu\alpha}^{\;\;\;\;\lambda}\hat{\nabla}_{\nu}(\sqrt{-g}\Delta_{\lambda}^{\;\;\;\mu\nu})\Big]
\end{gather}
So, since the matter action is invariant under diffeomorphisms we have the following conservation law
\beq
\sqrt{-g}(2 \tilde{\nabla}_{\mu}T^{\mu}_{\;\;\alpha}-\Delta^{\lambda\mu\nu}R_{\lambda\mu\nu\alpha})+\hat{\nabla}_{\mu}\hat{\nabla}_{\nu}(\sqrt{-g}\Delta_{\alpha}^{\;\;\mu\nu})+2S_{\mu\alpha}^{\;\;\;\;\lambda}\hat{\nabla}_{\nu}(\sqrt{-g}\Delta_{\lambda}^{\;\;\;\mu\nu})=0 \label{ccc}
\eeq
where we have defined as usual the energy-momentum tensor
\beq
T^{\mu\nu}:=+\frac{2}{\sqrt{-g}}\frac{\delta ( \sqrt{-g}\mathcal{L}_{M})}{\delta g_{\mu\nu}}
\eeq
The hypermomentum
\beq
\Delta_{\lambda}^{\;\;\;\mu\nu}:=-\frac{2}{\sqrt{-g}}\frac{\delta ( \sqrt{-g}\mathcal{L}_{M})}{\delta \Gamma^{\lambda}_{\;\;\;\mu\nu}}
\eeq
and the modified covariant derivative
\beq
\hat{\nabla}_{\mu}:=2S_{\mu}-\nabla_{\mu}
\eeq
Then, solving for $\hat{\nabla}_{\nu}(\sqrt{-g}\Delta_{\lambda}^{\;\;\mu\nu})$ from
	\beq
t^{\mu}_{\;\;\lambda}
= T^{\mu}_{\;\;\lambda}-\frac{1}{2 \sqrt{-g}}\hat{\nabla}_{\nu}(\sqrt{-g}\Delta_{\lambda}^{\;\;\mu\nu}) \label{cc1}
\eeq
and substituting it back to ($\ref{ccc}$) we recover the second conservation law
\beq
\frac{1}{\sqrt{-g}}\hat{\nabla}_{\mu}(\sqrt{-g}t^{\mu}_{\;\;\alpha})=\frac{1}{2} \Delta^{\lambda\mu\nu}R_{\lambda\mu\nu\alpha}+\frac{1}{2}Q_{\alpha\mu\nu}T^{\mu\nu}+2 S_{\alpha\mu\nu}t^{\mu\nu} \label{cc2}
\eeq
showing that  the sets $(\ref{ccc})$, $(\ref{cc1})$ and $(\ref{cc1})$, $(\ref{cc2})$  are indeed equivalent.

\section{ Scalar field coupled to the connection}

As a trivial example of an effective Perfect Cosmological Hyperfluid, endowed with specific equations of state among its variables, we will consider a scalar field $\varphi$ that is coupled to the connection. The most straightforward couplings would be between the derivatives of the scalar field and  torsion and non-metricity vectors, such that  the total matter action would read\footnote{A similar action was considered in \cite{shimada2019metric} in order to study inflation-like scenarios in MAG.}
\beq
S_{M}[g,\Gamma, \varphi]=\int d^{n}x \sqrt{-g}\left[ -\frac{1}{2}g^{\mu\nu}\partial_{\mu}\varphi \partial_{\nu}\varphi-V(\varphi)+(a_{1}S_{\mu}+a_{2}Q_{\mu}+a_{3} \tilde{Q}_{\mu})\partial^{\mu}\varphi \right]
\eeq
where  $a_{i}$  with $i=1,2,3$ are coupling constants. The associated hypermomentum can be trivially computed
\beq
\Delta_{\lambda}^{\;\;\mu\nu}=-(a_{1}+2 a_{3})\delta_{\lambda}^{\nu}\partial^{\mu}\varphi +(a_{1}-4 a_{2})\delta_{\lambda}^{\mu}\partial^{\nu}\varphi -2 a_{3}g^{\mu\nu}\partial_{\lambda}\varphi
\eeq
Then, lowering all indices and considering an FLRW spacetime so that $\partial_{\mu}\varphi=\delta_{\mu}^{0}\dot{\varphi}=-\dot{\varphi}u_{\mu}$ holds true, the above recasts to
\beq
\Delta_{\alpha\mu\nu}=(a_{1}+2 a_{3})g_{\alpha\nu}u_{\mu}\dot{\varphi} +(4 a_{2}-a_{1})g_{\alpha\mu} u_{\nu} \dot{\varphi} +2 a_{3}g_{\mu\nu}u_{\alpha}\dot{\varphi}
\eeq
In addition, using the definition of the projection operator and placing the above equation against ($\ref{Dform}$) we find the relations
\begin{gather}
\phi=(4 a_{2}-a_{1})\dot{\varphi} \;\;, \;\;\; \chi=(a_{1}+2 a_{3})\dot{\varphi} \;\;, \;\;\; \psi=2 a_{3}\dot{\varphi} 
 \omega=-4(a_{2}+a_{3})\dot{\varphi}\;\;,\;\;\; \zeta=0
\end{gather}
which serve as effective equations of state for the hyperfluid variables. This can be seen clearly by considering one of the above fields as  a reference field. Indeed, if we take for instance $\phi$ as our reference field, the rest of the hypermomentum variables are expressed as
\beq
\chi=w_{1} \phi \;\;, \;\;\; \psi=w_{2}\phi \;\;, \;\;\; \omega=w_{3} \phi \;\;, \;\;\; \zeta=0
\eeq
where
\beq
(w_{1},w_{2},w_{3})=\frac{1}{(4 a_{2}-a_{1})}\Big( a_{1}+2 a_{3} \;, \; 2 a_{3}\;,\; -4 a_{2}-4 a_{3} \Big)  
\eeq
are the 'barotropic' indices of the hyperfluid in close analogy with the barotropic index $w$ in the case of the perfect fluid (recall the relation $p=w \rho$).
Note the similarity with the perfect fluid inflaton field, where in the slow roll regime ($\dot{\varphi}^{2}\approx 0$) it holds that
\beq
\rho \approx V(\varphi) \;\;, \;\;\; p\approx -V(\varphi)
\eeq
and one gets the equation of state $p\approx -\rho$.

	\bibliographystyle{unsrt}
	\bibliography{ref}

	\end{document}